\documentclass[11pt,preprint]{aastex} 

\usepackage{natbib}
\received{}
\accepted{}
\shorttitle{$H_0$ from a Refurbished Distance Ladder}
\shortauthors{Riess et al.}

\newcommand{\bq}{\begin{equation}} 
\newcommand{\eq}{\end{equation}}   
 
\newcommand{\ho}{$74.8 \pm 3.1$ km s$^{-1}$ Mpc$^{-1}$}  
\newcommand{\honosys}{$74.8 \pm 3.0$ km s$^{-1}$ Mpc$^{-1}$}  
\newcommand{\holmcnosys}{$71.3 \pm 3.8$ km s$^{-1}$ Mpc$^{-1}$}

\newcommand{\homwnosys}{$75.7 \pm 2.6$ km s$^{-1}$ Mpc$^{-1}$}
\newcommand{\hobothnosys}{$74.5 \pm 2.3$ km s$^{-1}$ Mpc$^{-1}$}
\newcommand{\homwlmc}{$74.4 \pm 2.5$ km s$^{-1}$ Mpc$^{-1}$}
\newcommand{\hoall}{$73.8 \pm 2.1$ km s$^{-1}$ Mpc$^{-1}$}  %except without systematic error
\newcommand{\hoalle}{$73.8 \pm 2.3$ km s$^{-1}$ Mpc$^{-1}$}  % no sys, error from just 2
\newcommand{\hofin}{$73.8 \pm 2.4$ km s$^{-1}$ Mpc$^{-1}$} % above with sys
\newcommand{\uncfin}{3.3\% }  % final %, just 2
\newcommand{\unc}{4.1\% }  % complete
\newcommand{\uncbothnosys}{3.0\% }  % statistical
\newcommand{\uncsnosys}{4.0\% }  % statistical
\newcommand{\uncmwlmc}{3.4\% }  % complete
\newcommand{\uncallnosys}{2.9\% }  % statistical
\newcommand{\beq}{\begin{equation}}
\newcommand{\eeq}{\end{equation}}
\newcommand{\beqa}{\begin{eqnarray}}
\newcommand{\eeqa}{\end{eqnarray}}

\newcommand{\PL}{$P$--$L$\ }

\begin{document} 

\title{A 3\% Solution: Determination of the Hubble Constant with the 
Hubble Space Telescope and Wide Field Camera 3 \altaffilmark{1}}

\vspace*{0.3cm}

%{\it ApJ, 730, 119, 2011 }

\vspace*{0.3cm}

\author{Adam G. Riess\altaffilmark{2,3}, Lucas Macri\altaffilmark{4},
Stefano Casertano\altaffilmark{3},  Hubert Lampeitl\altaffilmark{5},
Henry C. Ferguson\altaffilmark{3}, Alexei V. Filippenko\altaffilmark{6},
Saurabh W. Jha\altaffilmark{7}, Weidong Li\altaffilmark{6}, and
Ryan Chornock\altaffilmark{8}}

\altaffiltext{1}{Based on observations with the NASA/ESA {\it Hubble Space
  Telescope}, obtained at the Space Telescope Science Institute, which is
  operated by AURA, Inc., under NASA contract NAS 5-26555.}
\altaffiltext{2}{Department of Physics and Astronomy, Johns Hopkins
  University, Baltimore, MD 21218.}
\altaffiltext{3}{Space Telescope Science Institute, 3700 San Martin
  Drive, Baltimore, MD 21218; ariess@stsci.edu.}
\altaffiltext{4}{George P. and Cynthia Woods Mitchell Institute for
  Fundamental Physics and Astronomy,\\Department of Physics \&
  Astronomy, Texas A\&M University, 4242 TAMU, \\ College Station, TX
  77843-4242.}
\altaffiltext{5}{ Institute of Cosmology and Gravitation, University
  of Portsmouth, Portsmouth, PO1 3FX, UK}
\altaffiltext{6}{Department of Astronomy, University of California,
  Berkeley, CA 94720-3411.}
\altaffiltext{7}{Department of Physics and Astronomy, Rutgers University,
  136 Frelinghuysen Road, Piscataway, NJ 08854.}
\altaffiltext{8}{Harvard/Smithsonian Center for Astrophysics, 60 Garden 
  St. Cambridge, MA 02138.}

%\author{Adam G. Riess\altaffilmark{2}}
%\affil{Department of Physics and Astronomy, Johns Hopkins University}
%\affil{Baltimore, MD 21218}
%\email{ariess@stsci.edu}

%\author{Lucas Macri}
%\affil{George P. and Cynthia Woods Mitchell Institute for Fundamental Physics 
% and Astronomy}
%\affil{Department of Physics \& Astronomy, Texas A\&M University}
%\affil{4242 TAMU, College Station, TX 77843-4242}
%\email{lmacri@tamu.edu}

%\author{Stefano Casertano}
%\affil{Space Telescope Science Institute}
%\affil{3700 San Martin Drive, Baltimore, MD 21218}
%\email{stefano@stsci.edu}

%\author{Hubert Lampeitl}
%\affil{Space Telescope Science Institute}
%\affil{3700 San Martin Drive, Baltimore, MD 21218}
%\email{stefano@stsci.edu}

%\author{Henry C. Ferguson}
%\affil{Space Telescope Science Institute}
%\affil{3700 San Martin Drive, Baltimore, MD 21218}
%\email{ferguson@stsci.edu}

%\author{Alexei V. Filippenko}
%\affil{Department of Astronomy, University of California at Berkeley}
%\affil{601 Campbell Hall, Berkeley, CA 94720-3411} 

%\author{Saurabh W. Jha}
%\affil{Department of Physics and Astronomy, Rutgers University}
%\affil{136 Frelinghuysen Road, Piscataway, NJ 08854}

%\author{Weidong Li}
%\affil{Department of Astronomy, University of California at Berkeley}
%\affil{601 Campbell Hall, Berkeley, CA 94720-3411} 

%\author{Ryan Chornock}
%\affil{Department of Astronomy, University of California at Berkeley}
%\affil{601 Campbell Hall, Berkeley, CA 94720-3411} 

\begin{abstract} 
  We use the Wide Field Camera 3 (WFC3) on the {\it Hubble Space Telescope
  (HST)} to determine the Hubble constant from optical and infrared
  observations of over 600 Cepheid variables in the host galaxies of 8 recent
  Type Ia supernovae (SNe~Ia), which provide the calibration for a
  magnitude-redshift relation based on 240 SNe~Ia.  Increased precision over
  past measurements of the Hubble constant comes from five improvements: (1)
  more than doubling the number of infrared observations of Cepheids in the
  nearby SN hosts; (2) increasing the sample size of ideal SN~Ia calibrators
  from six to eight with the addition of SN 2007af and SN 2007sr; (3) increasing
  by 20\% the number of Cepheids with infrared observations in the megamaser
  host NGC\,4258; (4) reducing the difference in the mean metallicity of the
  Cepheid comparison samples between NGC\,4258 and the SN hosts from $\Delta
  {\rm log\,[O/H]} = 0.08$ to 0.05; and (5) calibrating all optical Cepheid
  colors with a single camera, WFC3, to remove cross-instrument zeropoint
  errors. The result is a reduction in the uncertainty in $H_0$ due to steps
  beyond the first rung of the distance ladder from 3.5\% to 2.3\%.  The
  measurement of $H_0$ via the geometric distance to NGC\,4258 is \ho, a \unc\
  measurement including systematic uncertainties.  Better precision independent
  of the distance to NGC\,4258 comes from the use of two alternative Cepheid
  absolute calibrations: (1) 13 Milky Way Cepheids with trigonometric parallaxes
  measured with {\it HST}/FGS and {\it Hipparcos}, and (2) 92 Cepheids in the
  Large Magellanic Cloud, for which multiple accurate and precise eclipsing
  binary distances are available yielding \homwlmc, a \uncmwlmc\ uncertainty
  including systematics.  Our best estimate uses all three calibrations but a
  larger uncertainty afforded from any two: $H_0 =$ \hofin\ including systematic
  errors, corresponding to a \uncfin uncertainty. The improved measurement of
  $H_0$, when combined with the Wilkinson Microwave Anisotropy Probe (WMAP)
  7-year data, results in an improved constraint on the equation-of-state
  parameter of dark energy of $w = -1.08 \pm 0.10$.  It also rules out the best
  fitting gigaparsec-scale void models, posited as an alternative to dark
  energy.  The combined $H_0$ + WMAP results also have implications for the
  number of relativistic particle species in the early Universe, yielding
  $N_{\rm eff}=4.2 \pm 0.7$, an excess for the value expected from the three
  known neutrino flavors, though not with high significance.  The distance
  ladder used here for the determination of $H_0$ does not yet appear to be
  limited by systematic errors, suggesting that further improvements in
  precision approaching 1\% may be feasible.
\end{abstract} 


\section{Introduction} 

Measurements of the expansion history, $H(z)$, from Type Ia supernovae (SNe~Ia)
provide crucial, empirical constraints to help guide the emerging cosmological
model. While high-redshift SNe~Ia reveal that the Universe is now accelerating
\citep{riess98,perlmutter99}, nearby ones provide the most precise measurements
of the present expansion rate, $H_0$.

Recently, high-redshift measurements from the cosmic microwave background
radiation (CMB), baryon acoustic oscillations (BAO), and SNe~Ia have been used
in concert with an assumed cosmological model to {\it predict} the value of
$H_0$ \citep[e.g.,][]{komatsu11}.  They are not, however, a substitute for its
{\it measurement} in the local Universe.  Such forecasts of $H_0$ from the
high-redshift Universe also make specific assumptions about unsettled questions:
the nature of dark energy, the global geometry of space, and the basic
properties of neutrinos (number and mass).  Instead, we can gain insights into
these unknowns from a precise, local measurement of $H_0$.  The most precise
measurements of $H_0$ have come from distance ladders which calibrate the
luminosities of nearby SNe~Ia through {\it Hubble Space Telescope (HST)}
observations of Cepheids in their host galaxies (see \citealt {freedman10} for a
review).
                                                                                
In the early Cycles of {\it HST}, the SN~Ia {\it HST} Calibration Program
\citep[hereafter SST]{sandage06} and the {\it HST} Key Project \citep[hereafter
KP]{freedman01} each calibrated $H_0$ via Cepheids and SNe~Ia using the Wide
Field/Planetary Camera 2 (WFPC2) and the Large Magellanic Cloud (LMC) as the
first rung on their distance ladder. Unfortunately, the LMC was not an ideal
anchor for the cosmic ladder because its distance was constrained to only
5--10\% \citep{gibson00a}, its Cepheids (observed from the ground) are of
shorter mean period ($\Delta <P> \approx 25$~d), and lower metallicity ($\Delta
{\rm [O/H]} = 0.4$) than those of the spiral galaxies hosting nearby
SNe~Ia. These differences and uncertainties between ground-based and space-based
photometric zeropoints introduced a 7\% systematic error in the determinations
of $H_0$ obtained by those teams (see \S 4). Additional uncertainty arose from
the unreliability of the measurements from several of the SNe~Ia selected by
SST, which were {\it photographically} observed, highly reddened, atypical, or
discovered after peak brightness.  Only three SNe~Ia (SNe 1990N, 1981B, and
1998aq) from the SST sample lacked these shortcomings, defining only a small set
of nearby SNe suitable to calibrate $H_0$.  Despite careful work, the teams'
estimates of $H_0$, each with an uncertainty of $\sim 10$\%, differed from each
other's by 20\%, due to the aforementioned systematic errors. Additional
progress required rebuilding the distance ladder to address these systematic
errors.

The installation of the Advanced Camera for Surveys (ACS) extended the
range of {\it HST} for observing Cepheids, reduced their crowding with
finer pixel sampling, and increased their rate of discovery by
doubling the field of view.  In Cycle 11, members of our team began
using ACS to measure Cepheids at optical wavelengths in the hosts of
more modern SNe~Ia (SN 1994ae by \citealt{riess05}; SN 1995al and SN
2002fk by \citealt{riess09b}) and in a more ideal anchor galaxy
\citep[NGC\,4258 by][]{macri06}.

In {\it HST} Cycle 15, we began the ``Supernovae and $H_0$ for the
Equation of State'' (SH0ES) project to measure $H_0$ to better than
5\% precision by addressing the largest remaining sources of
systematic error.  The SH0ES program constructed a refurbished
distance ladder from high-quality light curves of SNe~Ia, a geometric
distance to NGC\,4258 determined through radio (very long baseline
interferometry; VLBI) observations of megamasers, and Cepheid
variables observed with {\it HST} in NGC\,4258 and in the hosts of
recent SNe~Ia.  The reduction in systematic errors came from additional observations of NGC\,4258 and from our use of purely
{\it differential} measurements of the fluxes of Cepheids with similar
metallicities and periods throughout all galaxies in our sample. The
latter rendered our distance scale insensitive to possible changes in
Cepheid luminosities as a function of metallicity or to putative
changes in the slope of the period-luminosity relations from galaxy to
galaxy. We measured $H_0$ to 4.7\% precision \citep[hereafter
R09]{riess09a}, a factor of two better than previous measurements
with {\it HST} and WFPC2.  An alternate analysis using the
\citet{benedict07} parallax measurements of Milky Way Cepheids in lieu
of the megamaser distance to NGC\,4258 showed good agreement, with comparable
5.5\% precision.

This result formed a triumvirate of constraints in the
Wilkinson Microwave Anisotropy Probe (WMAP) 7-year analysis (i.e., BAO
+ $H_0$ + WMAP-7yr) which were selected as the combination most insensitive to systematic
errors with which to constrain the cosmological parameters
\citep{komatsu11}.  Together with the WMAP constraint on $\Omega_M
h^2$, this measurement of $H_0$ provides a constraint on the nature of dark energy, 
$w = -1.12 \pm 0.12$ \citep[R09 and][]{komatsu11},
which is comparable to but independent of the use of high-redshift SNe~Ia.
It also improves constraints on the
properties of the elusive neutrinos, such as the sum of their masses and the
number of species \citep{komatsu11}.

In {\it HST} Cycle 17 we used the newly installed Wide Field Camera 3
(WFC3) to increase the sample sizes of both the Cepheids and the SN~Ia
calibrators along the ladder used by SH0ES to determine $H_0$.  The
near-infrared (IR) channel of WFC3 provides an order of magnitude
improvement in efficiency for follow-up observations of Cepheids over
the Near-Infrared Camera and Multi-Object Spectrograph (NICMOS), while
the finer pixel scale of the visible channel (relative to ACS) is
valuable for reducing the effects of crowding when searching for
Cepheids. We present these new observations in \S2, the
redetermination of $H_0$ in \S3, and an analysis of the error budget
including systematics in \S4. In \S5, we address the use of this new
measurement along with external datasets to constrain properties of
dark energy and neutrinos.

\section{WFC3 Observations of Cepheids in the SH0ES Program}

The SH0ES program was developed to improve upon the calibration of the
luminosity of SNe~Ia in order to better measure the Hubble constant.  To ensure
a reliable calibration sample we selected SNe~Ia having the following qualities:
(1) modern photometric data (i.e., photoelectric or CCD), (2) observed before
maximum brightness, (3) low reddening (implying $A_V < 0.5$ mag), (4)
spectroscopically normal, and (5) optical {\it HST}-based observations
of Cepheids in its host galaxy.  In addition to providing robust distance
measures, these qualities are crucial for producing a calibration sample which
is a good facsimile of the SN~Ia sample they are used to calibrate --- i.e.,
those defining the modern SN~Ia magnitude-redshift relation at $0.01 < z < 0.1$
\citep[e.g.,][]{hicken09a}.

In {\it HST} Cycles 16 and 17, we used WFC3, ACS, and WFPC2 to discover Cepheids
in two new SN~Ia hosts: NGC\,5584 \citep[host of SN 2007af;][]{macri11a} and
NGC\,4038/9 \citep[``the Antennae,'' host of SN 2007sr;][]{macri11b} whose light
curves were presented in \citet{hicken09a}\footnote{ We augmented the Hicken et
  al light curve of SN 2007sr with 3 pre-discovery V-band observations from the
  All-Sky Automated Survey (ASAS) extending to pre-maximum; MJD,mag,error
  triplets are (4441.85,13.44,0.12), (4448.86,12.65,0.05),(4452.85,12.73,0.08)}.
We also employed the optical channel of WFC3 to {\it reobserve all previous
SN~Ia hosts in the calibration sample and NGC\,4258}.  This provided for the
first time a calibration of all Cepheid optical and infrared photometry using
the same zeropoints. In the case of some hosts, the additional epoch (obtained
well after the prior ones) allowed us to discover previously unidentified,
longer period ($P > 60$~d) variables. We also used these observations to search
for additional Cepheids in the hosts which previously had the smallest numbers
of Cepheids: NGC\,3021, NGC\,3982, NGC\,4536, and NGC\,4639
\citep{macri11c}. The new observations, together with those from
\citet{riess09b}, \citet{saha96,saha97,saha01}, \citet{gibson00b},
\citet{stetson01}, and \citet{macri06}, provide the position, period, and phase
of 730 Cepheids in 8 hosts with reliable SN~Ia data as well as NGC\,4258.  The
Cepheids in each host were typically imaged on 14 epochs in $F555W$ and 2--5
epochs with $F814W$ (except for NGC\,4258, which has 12 epochs of $F814W$
data). An illustration of the entire dataset used to observe the Cepheids is
shown in Figure 1.  Having previously determined the positions, periods, and
optical magnitudes of these Cepheids, it is highly advantageous to observe their
near-IR magnitudes with a single photometric system in order to (1) reduce the
differential extinction by a factor of five over visual bands, (2) reduce the
possible dependence of Cepheid luminosities on chemical composition
\citep{marconi05}, and (3) negate zeropoint errors.  This was previously done
with NICMOS on {\it HST} in Cycle 15 by R09 for a subset of these Cepheids.

The near-IR channel on WFC3 provides a tremendous gain over
NICMOS for the study of extragalactic Cepheids. Photometry of
comparable signal-to-noise ratio can be obtained in a quarter of the
exposure time.  More significant for Cepheid follow-up observations is
the factor of 40 increase in area of WFC3-IR over NICMOS/Camera~2
(NIC2), the channel which offered the best compromise between area and
uniform pixel sampling.  The one advantage of NIC2 over WFC3-IR is
better sampling of the point-spread function (PSF); the $0\farcs13$
pixels of WFC3-IR undersample the {\it HST} PSF by a factor of 1.6 at
1.6~$\mu$m. However, the finer sampling of NIC2 is largely offset by
the numerous photometric anomalies unique to that camera, whose
subsequent correction leads to correlated noise among neighboring
pixels which reduces the independence of the NICMOS pixel sampling.
In contrast, the detector of WFC3-IR is much better behaved and
pixel sampling noise can be mitigated with dithering.

\subsection{WFC3 Data Reduction}

Each host galaxy was observed for 2--7~ks with individual exposures
400--700~s in length, using integer and half-pixel dithering between
exposures to improve sampling of the PSF (see Table 1).  The WFC3
images of the two new hosts, NGC\,5584 and NGC\,4038, are shown in
Figures 2 and 3. Figure 4 shows an example of a host previously
observed with NIC2 by R09 and with WFC3-IR in this study.

We developed an automated pipeline to calibrate the raw WFC3 $F160W$
frames. The first step was to pass the data through the
STScI-supported {\it calwf3} pipeline in the {\it STSDAS} suite of
routines in {\it PyRAF} to remove the bias and dark current, reject
cosmic rays through up-the-ramp sampling, and flat-field the data.  A
small correction to the standard flat-field frame was used to correct
the WFC3 ``blobs,'' which are 10\% depressions in flux covering $\sim
1$\% of the area due to spotting on the WFC3 Channel Select Mechanism
(CSM).  Next, we used {\it multidrizzle} to combine the exposures from
each visit into a master image, resampling onto a finer pixel scale
while correcting for the known geometric distortions in the camera. We
utilized a final pixel scale of $0\farcs08$ per pixel and an
input-to-output fraction of 0.6.

We identified the positions of Cepheids in the WFC3-IR
images by deriving geometric transformations from the $F814W$ images
to those in $F160W$, successively matching fainter sources to improve
the registration.  This procedure empirically determined the
difference in plate scale between ACS-WFC, WFPC2, WFC3-UVIS, and
WFC3-IR.  We typically identified more than 100 sources in common,
resulting in an uncertainty in the mean position of each Cepheid below
0.03 pixels ($<2.4$ milliarcsec).

We carried out the photometry of Cepheids using the algorithms
developed by R09; they employ PSF fitting to model the crowded regions
around Cepheids, fixing their positions to those derived from optical
data and using artificial-star tests to determine photometric errors
and crowding biases. As an example, we show in Figure 5 the {\it HST}
optical image, near-IR image, model, and residuals for 8 typical
Cepheids spanning a wide range of periods in the SN host NGC\,5584.
We used the same approach to determine zeropoints as in R09 from the Persson et al. (1998) standard star
P330E. 

\begin{table}[h]
\tablenum{2}
\begin{small}
\begin{center}
\vspace{0.4cm}
\begin{tabular}{lll}
\multicolumn{3}{c}{Table 1: Hosts Observed with WFC3-IR $F160W$ by GO-11570} \\
\hline
\hline
Host & SN~Ia &  Exp. time (s)   \\
\hline
NGC\,4536 & SN 1981B     &  2564 \\
NGC\,4639 & SN 1990N     &  5377 \\
NGC\,3982 & SN 1998aq    &  4016 \\
NGC\,3370 & SN 1994ae    &  4374 \\
NGC\,3021 & SN 1995al    &  4424 \\
NGC\,1309 & SN 2002fk    &  6988 \\
NGC\,4038/9 & SN 2007sr & 6794$^a$ \\
NGC\,5584 & SN 2007af & 4926 \\
NGC\,4258 & ------------ &  2011$^b$ \\
\hline
\hline
 \multicolumn{3}{l}{$^a$Data in GO-11577} \\
 \multicolumn{3}{l}{$^b$Depth per pointing; galaxy covered in 16 pointing mosaic} \\
\end{tabular}
\end{center}
\end{small} 
\end{table}

Due to the low amplitudes of their near-IR light curves ($< 0.3$ mag),
Cepheid magnitudes determined at random phases provide nearly the same precision as mean fluxes for
determining the intercept
of their \PL relations \citep{madore91}.\footnote{In R09 we
  corrected the measured NICMOS $F160W$ magnitude to the mean-phase
  magnitude using the Cepheid phase, period, and amplitude from the
  optical data, the dates of the NICMOS observations, and the Fourier
  components of \citet{soszynski05} which quantify the relations
  between Cepheid light curves in the optical and near-IR.  However,
  these phase corrections of $\sigma \sim 0.1$ mag were found to be
  insignificant in the subsequent analysis, since the dispersion of
  the observed \PL relations is $\sigma \sim 0.3$ mag.  Here we have not
  attempted such corrections because the Cepheid phases at the time of
  the WFC3 $F160W$ observations were too poorly constrained to allow
  for a significant correction.}

Since we had previously observed with NIC2 many of the Cepheids now
observed with WFC3-IR, we can directly compare their $F160W$
photometry on these two systems.  Figure 6 shows the magnitude
differences for the Cepheids utilized in the \PL relations in both R09
and in \S 2.2.  The mean difference is $0.036 \pm 0.027$~mag (in the
sense that photometry with WFC3 is brighter), with no apparent
dependence on Cepheid brightness.  While the difference in photometry
between instruments may include differences in system zeropoints, the
subsequent determination of $H_0$ via Cepheids observed with a single
instrument in the SN~Ia hosts and in NGC\,4258 will be independent of
instrument zeropoints.  Thus, for the determination of $H_0$ it is
more relevant to calculate the change in magnitudes between Cepheids
in NGC\,4258 and the SN hosts between WFC3 and NIC2; the measurement
of this change is $0.019 \pm 0.054$~mag.

Table 2 contains relevant parameters for each Cepheid observed with WFC3
$F160W$.  The first 8 columns give the Cepheid's host, position, identification
number (from \citealt{macri06}, \citealt{riess09b}, \citealt{macri11a, macri11b,
  macri11c}), period, mean $V-I$ color, WFC3 $F160W$ magnitude, and the
magnitude uncertainty.  Column 9 contains the displacement of the flux centroid
in the near-IR data relative to the optical Cepheid position, expressed in units
of pixels (1 pixel $ = 0\farcs08$), a quantity used to refine the determination
of the crowding bias. Column 10 gives the photometric crowding bias determined
using the artificial-star tests for each Cepheid's environment (see \S 2.3 of
R09) and the displacement tabulated in the previous column; this correction has
already been applied to the magnitudes listed in Column 7.  Column 11 contains
the root-mean square of the residual image, weighted by the inverse distance
from the Cepheid position, useful for determining the quality of the
crowded-scene fit. Column 12 contains the metallicity parameter, 12 + log\,[O/H]
\citep{zaritsky94} derived from the deprojected galactocentric
radii of each Cepheid and the abundance gradient of its host.\footnote{These
  gradients were published in R09 for NGC\,4258 and the previously observed six
  SN hosts; the values for the new hosts, following the same convention as Table
  12 of that paper, are 12 + log\,[O/H] $= 8.981 - 0.064(x-30\arcsec)/10\arcsec$
  for NGC\,5584 and 12 + log\,[O/H] $= 9.129 - 0.043(x-30\arcsec)/10\arcsec$ for
  NGC\,4038/9.}  Column 13 contains a rejection flag used for the \PL relations.

\subsection{Near-Infrared Cepheid Relations}

The nine individual \PL relations measured with WFC3 $F160W$ and fit with a
common slope are shown in Figure 7. Intercepts relative to NGC\,4258 are given
in Table 3 and compared in Figure 8 to the SN distances.  While 636 Cepheids previously identified at optical
wavelengths were measurable\footnote{Cepheids were considered to be measured if
  our software reported a possible magnitude for the source with an uncertainty
  less than 0.7 mag, a model residual with rms better than $3 \sigma $ from the
  distribution of all model residuals, and a crowding correction less than 1.5
  magnitudes.  While these thresholds are somewhat arbitrary, they are
  sufficient to remove catastrophic failures in convergence on the measurement
  of the photometry for a source.}  in the WFC3-IR $F160W$ images, $\sim 20$\%
appeared as outliers in the IR \PL relations.  This is not surprising, as we
expect outliers to occur from (1) a complete blend with a bright, red source
such as a red giant, (2) a poor model reconstruction of a crowded group when
the Cepheid is a small component of the group's flux, (3) objects misidentified
as classical Cepheids in the optical (e.g., blended Type II Cepheids), and (4)
Cepheids with the wrong period (aliasing or incomplete sampling of a single
cycle).  As expected, the outlier fraction is greater in WFC3 images than in
NIC2 ones because the former contain a larger fraction of Cepheids from crowded
regions (such as the nucleus) which yield more outliers and were intentionally
avoided in the small, selective NIC2 pointings (see Figure 4).

As in R09, we eliminated outliers $> 0.75$ mag or $> 2.5\sigma$ (following
Chauvenet's criterion) from an initial fit of the \PL relations, refitted the
relations and repeated these tests for outliers until convergence.  This
resulted in a reduction of the sample to 484 objects; the next section
considers the effect of this rejection on the determination of $H_0$ and an
alternative method for contending with outliers.

\section{Measuring the Hubble Constant}

The determination of the Hubble constant follows from the relations
given in \S 3 of R09.  To summarize, we perform a single, simultaneous
fit to all Cepheid and SN~Ia data to minimize the $\chi^2$ statistic
and measure the parameters of the distance ladder.  We express the
{\it j}th Cepheid magnitude in the {\it i}th host as
  
\bq m_{W,i,j}=(\mu_{0,i}-\mu_{0,4258})+zp_{W,4258}+b_W \ {\rm log} \
P_{i,j}+Z_W \ \Delta {\rm log\,[O/H]}_{i,j}, \eq

\noindent where the ``Wesenheit reddening-free'' mean magnitude
\citep{madore82} is given as

\bq m_{W,i,j} =m_{H,i,j} - R(m_{V,i,j} - m_{I,i,j}), \eq 

\noindent and $R \equiv A_H / (A_V - A_I)$.  The Cepheid parameters with $i,j$
subscripts are given in Table 2.  For a \citet{cardelli89} reddening law, a
Galactic-like value of $R_V = 3.1$, and the $H$ band corresponding to the WFC3
$F160W$ band, we have $R = 0.410$.  In the next section we consider the
sensitivity of $H_0$ to the value of $R_V$.

We determine the values of the nuisance parameters $b_W$ and $Z_W$ --- which
define the relation between Cepheid period, metallicity, and luminosity --- by
minimizing the $\chi^2$ for the global fit to all Cepheid data.  The
reddening-free distances, $\mu_{0,i}$, for the hosts relative to NGC\,4258 are
given by the fit parameters $\mu_{0,i}-\mu_{0,4258}$, while $zp_{4258}$ is the
intercept of the \PL relation simultaneously fit to the Cepheids of NGC\,4258.

The SN~Ia magnitudes in the SH0ES hosts are simultaneously expressed as

\bq m_{v,i}^0=( \mu_{0,i}-\mu_{0,4258})+m^0_{v,4258}. \eq
\noindent

\noindent where the value $m_{v,i}^0$ is the maximum-light apparent $V$-band
brightness of a SN~Ia in the {\it i}th host at the time of $B$-band peak
corrected to the fiducial color and luminosity.  This quantity is determined
for each SN~Ia from its multi-band light curves and a light-curve fitting
algorithm, either from the MLCS2k2 \citep{jha07} or the SALT-II \citep{guy05}
prescription (see \S 4.2 for further discussion).

A minor change from R09 is the inclusion of a recently identified, modest
relationship between host-galaxy mass and the calibrated SN~Ia
magnitude. Several studies \citep{hicken09b,kelly10,lampeitl10,sullivan10} have
shown the existence of a correlation between the corrected SN magnitude and the
mass of its host, with a value of 0.03 mag per dex in $M_{\rm stellar}$, in the
sense that more massive (and metal rich) hosts produce more luminous SNe. This
correlation has been independently detected using both low- and high-redshift
samples of SNe~Ia, as well as with multiple fitting algorithms.  The effect on
$H_0$ is quite small, a decrease of 0.75\%, due to the modest difference in
mean masses for the nearby hosts \citep[mean log $M_{\rm stellar} =
10.0$]{neill09} and for those that define the magnitude-redshift relation
\citep[mean log $M_{\rm stellar} = 10.5$]{sullivan10}. We include these
corrections based on host-galaxy mass in our present determination of
$m_{v,i}^0$, given in Table 3, normalizing to a fiducial host mass of log
$M_{\rm stellar} = 10.5$ as appropriate for the objects used to measure the
Hubble flow.

\begin{table}[h]
\tablenum{3}
\begin{small}
\begin{center}
\vspace{0.4cm}
\begin{tabular}{lllllll}
\multicolumn{6}{c}{Table 3: Distance Parameters} \\
\hline
\hline
Host & SN~Ia &  Filters   &  $m_{v,i}^0 +5a_v$  & $\sigma^a$ & $\mu_{0,i}-\mu_{0,4258}$ & $\mu_0$ best \\
\hline
n4536 & SN 1981B & $UBVR$ &  15.147  &  0.145  &  1.567  (0.0404)  & 30.91 (0.07) \\
n4639 & SN 1990N & $UBVRI$ &  16.040  &  0.111  &  2.383  (0.0630) & 31.67 (0.08) \\ 
n3370 & SN 1994ae & $UBVRI$ &  16.545  &  0.101  &  2.835  (0.0284) & 32.13  (0.07) \\
n3982 & SN 1998aq & $UBVRI$ &  15.953  &  0.091  &  2.475  (0.0460)  & 31.70 (0.08) \\
n3021 & SN 1995al & $UBVRI$ &  16.699  &  0.113  &  3.138  (0.0870)  & 32.27 (0.08) \\
n1309 & SN 2002fk & $BVRI$ &  16.768  &  0.103  &  3.276  (0.0491)  & 32.59 (0.09) \\
n5584 & SN 2007af & $BVRI$ &  16.274  &  0.122  &  2.461  (0.0401)  & 31.72 (0.07) \\
n4038 & SN 2007sr & $BVRI$ &  15.901  &  0.137  &  2.396  (0.0567)  & 31.66  (0.08) \\
Weighted Mean     &  -----    &  -----  &  -----  & 0.0417 & ------ (0.0133) & ----- \\
\hline
\hline
\multicolumn{6}{l}{$^a$For MLCS2k2, 0.08 mag added in quadrature to fitting error.} \\
\end{tabular}
\end{center}
\end{small} 
\end{table}

The simultaneous fit to all Cepheid and SN~Ia data via Equations (1) and (3)
results in the determination of $m^0_{v,4258}$, which is the expected
reddening-free, fiducial, peak magnitude of a SN~Ia appearing in NGC\,4258.
Lastly, the Hubble constant is determined from

\bq {\rm log} \ H_0={(m_{v,4258}^0-\mu_{0,4258})+5a_v+25 \over 5}. \eq

\noindent where $\mu_{4258,0}$ is the independent, geometric distance estimate
to NGC\,4258 obtained through VLBI observations of water megamasers orbiting
its central supermassive black hole
\citep{herrnstein99,humphreys05,argon07,humphreys08,greenhill09}.  The term
$a_v$ is the intercept of the SN~Ia magnitude-redshift relation, approximately
${\rm log} cz - 0.2m_v^0$ but given for an arbitrary expansion history as

\bq a_v = {\rm log} \ cz
\left\{ 1 + {1\over2}\left[1-q_0\right] {z} 
-{1\over6}\left[1-q_0-3q_0^2+j_0 \right] z^2
+ O(z^3) \right\} - 0.2m_v^0, \eq

\noindent measured from the set of SN~Ia ($z,m_v^0$) independent of any
absolute (i.e., luminosity or distance) scale.  As in R09, we determine $a_v$
from a Hubble diagram for 240 SNe~Ia from \citet{hicken09a} using MLCS2k2
\citep{jha07} or the SALT-II \citep{guy05} prescription to determine $m_v^0$.
Limiting the sample to $0.023 < z < 0.1$ (to avoid the possibility of a local,
coherent flow; $z$ is the redshift in the rest frame of the CMB) leaves 140
SNe~Ia.  (In the next section we consider a lower cut of $z>0.01$.)  Together
with the present acceleration $q_0=-0.55$ and prior deceleration $j_0=1$
\citep{riess07}, we find $a_v=0.698 \pm 0.00225$. Note that
\citet{ganeshalingam10} recently published light curves of large sample of
SNe~Ia from the Lick Observatory Supernova Search, but there is a large overlap
with those given by \citet{hicken09a}.  There are only 13 SNe~Ia at $z > 0.023$
not already included in our sample, and their inclusion would have a negligible
impact on the uncertainty in $a_v$, itself one of the smallest contributors to
the error in $H_0$.

The full statistical error in $H_0$ is the quadrature sum of the uncertainty in
the three {\it independent} terms in Equation (4): $\mu_{4258,0}$,
$m^0_{v,4258}$, and $5a_V$, where $\mu_{4258,0}$ is the geometric distance
estimate to NGC\,4258 by \citet{herrnstein99}, claimed by \citet{greenhill09}
to currently have a 3\% uncertainty.

\citet{hui06} point out that the peculiar velocities of SN~Ia hosts and their
correlations can produce an additional systematic error in the determination of
the SN~Ia $m$--$z$ relation used for cosmography.  However, by making use of a
map of the matter density field, it is possible to correct individual SN~Ia
redshifts for these peculiar flows \citep{riess97}. \citet{neill07} made use of
the IRAS PSCz density field \citep{branchini99} to determine the effect of the
density field on the low-redshift SN~Ia $m$--$z$ relation and its impact on the
equation-of-state parameter of dark energy, $w = P/(\rho c^2)$ (where $P$ is
pressure and $\rho c^2$ is energy density).  Using their results for a
light-to-matter bias parameter $\beta = 0.5$ and the dipole from \citet{pike05}
results in an increase of the mean velocity of the low-redshift sample and in
the Hubble constant by 0.4\% over the case of uncorrelated velocities at rest
with respect to the CMB.  We use a new estimate of this mean peculiar velocity
for the \citet{hicken09a} SN sample which is a slightly larger value of
0.5\%. We account for this and assume an uncertainty of 0.1\% resulting from a
$\pm 0.2$ error in the value of $\beta$.

The result is $H_0=$\honosys, a \uncsnosys\ measurement (top line, Table 4).
It is instructive to deconstruct the individual sources of uncertainty to
improve our insight into the measurement.  In principle, the covariance between
the data and parameters does not allow for an exact and independent allocation
of propagated error for each term toward the determination of $H_0$.  However,
in our case, the diagonal elements of the covariance matrices provide a very
good approximation to the individual components of error.  These are given in
Table 5 and shown in Figure 9 for past and present determinations of $H_0$.

A number of improvements since R09 are evident by comparing Columns 2 and 3 in
Table 5 and as shown in Figure 9. The uncertainty in $H_0$ from all of the terms independent of the
megamaser distance to NGC\,4258 is 2.3\%, 50\% smaller than these same terms in
R09, a result of the increased sample of Cepheids and SN calibrators. This term
includes uncertainties due to the form of the \PL relation, Cepheid metallicity
dependences, photometry bias, and zeropoints --- all of which were important
systematic uncertainties in past determinations of the Hubble constant
\citep[see Column 1, which contains the values from][]{freedman01}.  In this
analysis, as in R09, these uncertainties have been reduced by the collection of
samples of Cepheids whose measures (i.e., metallicity, periods, and photometric
systems) are a good match between NGC\,4258 and the SN hosts.  Here the
contribution from an unknown dependence of Cepheid luminosity on metallicity
has been furthered reduced by 40\% owing to a better match between the
metallicity of the Cepheid samples in NGC\,4258 and the expanded SN host
sample.  In R09, the mean metallicity of the NGC\,4258 Cepheid sample on the
ZHK abundance scale was 12 + log\,[O/H] =8.91, nearly the same as the present
mean of 8.90.  However, the mean metallicity of the Cepheid sample in the SN
hosts has risen from 8.81 to 8.85.  Some of this change can be attributed to
the inclusion of Cepheids closer to the nuclei of the hosts and some to the
inclusion of two new hosts, NGC\,5584 and NGC\,4038/9, with higher-than-average
metallicities.  The reduction in the mean abundance difference between
NGC\,4258 and the SN~Ia hosts from 0.077 to 0.045~dex results in a decrease of
the error propagated into $H_0$ from 1.1\% to 0.6\%.  A similar reduction is
seen with the use of Milky Way Cepheids whose mean metallicity of 8.9 is closer
to the mean of the new Cepheid sample in the SN hosts.  We consider an
alternative calibration of abundances from \citet{bresolin11} in \S 4.1.

\subsection {Buttressing the First Rung} 

In our present determination of $H_0$, the 3\% uncertainty in the distance to
NGC\,4258 claimed by \citet{greenhill09} is now greater than all other sources
combined (in quadrature). The next largest term, the uncertainty in mean
magnitude of the eight nearby SNe~Ia, is 1.9\%. To significantly improve upon
our determination of $H_0$, we would need an {\it independent} calibration of
the first rung of the distance ladder as good as or better than the
megamaser-based measurement to NGC\,4258 in terms of precision and reliability.
Independent calibration of the first rung is also valuable as an alternative to
NGC\,4258, should future analyses reveal previously unidentified systematic
errors affecting its distance measurement.

A powerful alternative has recently become available through high
signal-to-noise ratio measurements of the trigonometric parallaxes of Milky Way
Cepheids using the Fine Guidance Sensor (FGS) on {\it HST}.  \citet{benedict07}
reported parallax measurements for 10 Cepheids, with mean individual precision
of 8\% and an error in the mean of the sample of 2.5\%.  These were used in R09
as a test of the distance scale provided by NGC 4258, but the improvement in
precision beyond the first rung in the previous section suggests greater value
in their use to enhance the calibration of the first rung.

\citet{vanleeuwen07} reanalyzed {\it Hipparcos} observations and determined
independent parallax measurements for the same 10 Cepheids (albeit with half
the precision of {\it HST}/FGS) and for 3 additional Cepheids (excluding Polaris which is an overtone pulsator 
and whose estimated fundamental period is an outlier among the Cepheids pulsing in the fundamental mode).  The resulting
sample can be considered an independent anchor with a mean, nominal uncertainty
of just 1.7\%.  We use the combined parallaxes tabulated by
\citet{vanleeuwen07} and their $H$-band photometry as an alternative to the
Cepheid sample of NGC\,4258 by replacing Equation (1) for the Cepheids in the
hosts of SNe~Ia with

\bq m_{W,i,j}=\mu_{0,i}+M_{W,1}+b_W \ {\rm log} \ P_{i,j}+Z_W \ \Delta
{\rm log\,[O/H]}_{i,j}, \eq

\noindent where $M_{W,1}$ is the absolute Wesenheit magnitude for a Cepheid
with $P = 1$~d, and simultaneously fitting the Milky Way Cepheids with the
relation

\bq M_{W,i,j}=M_{W,1}+b_W \ {\rm log} \ P_{i,j}+Z_W \ \Delta 
{\rm log\,[O/H]}_{i,j} . \eq

\noindent Equation (3) for the SNe~Ia is replaced with 

\bq m_{v,i}^0=\mu_{0,i}-M_V^0. \eq

\noindent The determination of $M^0_V$ for SNe~Ia together with the previous
term $a_v$ then determines the Hubble constant,

\bq {\rm log} \ H_0={M_V^0+5a_v+25 \over 5}. \eq

Since the near-IR magnitudes of these Milky Way Cepheids have not been directly
measured with WFC3, the use of these variables requires an additional allowance
for possible differences in their photometry. These may arise from differences
in instrumental zeropoints, crowding, filter transmission functions, and
detector well depth at which the sources are measured together with an
uncertainty in detector linearity.  Analysis of the absolute photometry from
WFC3-IR \citep{kalirai09} and the ground system (e.g., 2MASS;
\citealt{skrutskie06}) claim absolute precision of 2\%--3\%.  We therefore
assume a {\it systematic} uncertainty in the relative magnitudes between {\it
  HST} WFC3 $F160W$ Cepheid photometry and the ground-based measurements of
Milky Way Cepheids on the $H$-band system of \citet{persson98} of 4\%.  This
reduces the effective precision of the parallax distance scale from 1.7\% to
2.6\%. The ground-based photometry of these Milky Way Cepheids is tabulated by
\citet{groenewegen99} and R09.  This systematic error is included in the global
fit as an additional calibration equation with uncertainty given in the error
correlation matrix.

When using the Milky Way Cepheids, we now include an external constraint on the
slope of the near-IR \PL relation.  No such constraint was necessary or even of
significant value in the previous section because the Cepheid periods in
NGC\,4258 (mean log\,$P = 1.51$) are so similar to those in the SN~Ia host
(mean log\,$P = 1.63$). In contrast, the mean period of the Milky Way sample
(mean log\,$P = 1.0$) is substantially lower, giving an unconstrained slope of
the \PL relation a greater and unrealistically large lever arm.  Following
analyses of optical and near-IR Cepheid data in the Milky Way \citep{fouque07}
and the LMC \citep{persson04,udalski99}, we adopt a conservative constraint on
the slope of the Wesenheit relation of $-3.3 \pm 0.1$~mag per dex in log\,$P$.

Using the Milky Way Cepheids instead of NGC\,4258 as the first rung of the
distance ladder gives \homwnosys, in good agreement with (and even greater
precision than) the NGC\,4258-based value.  However, an overall improvement in
precision is realized by the {\it simultaneous} use of both the Milky Way
parallaxes and the megamaser-based distance to NGC\,4258, yielding
\hobothnosys, a remarkably small uncertainty of \uncbothnosys.

Another opportunity to improve upon the first rung on the distance ladder comes
from the sample of $H$-band observations of Cepheids in the LMC by
\citet{persson04}.  Recent studies of detached eclipsing binaries (DEBs) by
different groups provide claims of a reliable and precise distance to the LMC.
\citet{guinan98}, \citet{fitzpatrick02}, and \citet{ribas02} studied three
B-type systems (HV2274, HV982, EROS1044) which lie close to the bar of the LMC
and therefore provide a good match to the Cepheid sample of \citet{persson04}.
The error-weighted mean of these is $49.2 \pm 1.6$ kpc\footnote{A fourth system
  \citep[HV5936,]{fitzpatrick03} is located several degrees away from the bar
  and yields a distance that is closer by 3$\sigma$. Additional lines of
  evidence presented in that paper suggest this system lies above the disk of
  the LMC, i.e., closer to the Galaxy.}.  \citet{pietrzynski09} analyzed
OGLE-051019.64-685812.3, an eclipsing binary system comprised of two giant
G-type stars also located near the barycenter of the LMC, and found a distance
of $50.2 \pm 1.3$ kpc.  The average result, 49.8~kpc, provides a good estimate
of the distance to the LMC\footnote{However, we note the analysis by
  \citet{schaefer08}, who suggests a level of agreement in recent distance
  estimates to the LMC which is too good to be consistent with statistics.}.
Here we retain the larger of the two previous uncertainties to estimate the
distance modulus as $18.486 \pm 0.065$~mag, or an effective error of $\pm
0.076$~mag when including the aforementioned 0.04~mag uncertainty between the
ground-based and {\it HST}-based near-IR photometric systems.  Using this
distance to the LMC and the Cepheid sample of \citet{persson04} yields
\holmcnosys, as seen in Table 4.

Combining all three first rungs (Milky Way, Large Magellanic Cloud, and
NGC\,4258) provides the most precise measurement of $H_0$: \hoall, a slightly
smaller uncertainty of \uncallnosys.  As expected, the use of all three anchors
for the distance ladder instead of just one has the largest impact on the
overall uncertainty, reducing the total contribution of the first rung to the
error from 3.3\% to 1.5\%.  However, a substantial penalty is paid for the
mixing of ground-based and space-based photometric systems and the resultant
uncertainties in Wesenheit or dereddened magnitudes, adding a 1.4\% error to
$H_0$ where for NGC\,4258 alone none pertained. Modest increases in error also
result from the larger difference in mean Cepheid metallicity (LMC) and period
(LMC and Milky Way).

Past determinations of the absolute distance scale have had a checkered
history, with revisions common.  Thus, it may be prudent to rely on no more
than any two of the three possible anchors of the distance scale in the
determination of $H_0$. The omission of NGC\,4258, Milky Way parallaxes, or the
LMC yields a precision in $H_0$ of 3.3\%, 3.2\%, and 3.0\%, respectively.  We
thus adopt as our best determination \hoalle, the measurement from all three
sources of the distance scale, but with the larger error associated from only
two independent origins of the distance scale.

Should future work revise the distance to any one of the absolute distance
scale determinations, we provide the following recalibration: $H_0$ decreases
by 0.25, 0.30, and 0.14 km s$^{-1}$ Mpc$^{-1}$ for each increase of 1\% in the
distance to either NGC\,4258, the Milky Way parallax scale, or the distance to
the LMC.

In the last column of Table 3 we also give the best estimate of the distance to
each host from the global fit to all first rungs, Cepheid and SN data.  These
are useful to compare to alternative methods of measuring distances to these
hosts or to place a sample of relative measures of SNe Ia distances onto an
absolute scale.  For example, there has been recent dissagreement on the
distance modulus of the Antennae (NGC 4038/9); \citet{saviane08} claim a value
of $\mu_0$=30.62 $\pm 0.17$~mag based on the apparent tip of the red giant
branch (TRGB), while \citet{schweizer08} obtain $\mu_0=31.74 \pm 0.27$~mag from
SN 2007sr and $\mu_0=31.51 \pm 0.16$~mag from a different determination of the
TRGB, in agreement with previous estimates by \citet{whitmore99} and
\citet{tonry00} based on flow-field models. Our result of $\mu_0=31.66 \pm
0.08$~mag (with the uncertainty based on the global fit) strongly favors the
``long'' distance to the Antennae.

Although we have been careful to propagate our statistical errors, as well as
past sources of systematic error such as metallicity dependence, system
zeropoint, and instrumental uncertainties, we now consider a broader range of
systematic uncertainties relating to alternative approaches to the analysis of
the data.

\section{Analysis Systematics}

In the preceding section we presented our preferred approach to analyzing the
Cepheid and SN~Ia data, incorporating uncertainties within the framework used
to model the data.  Here we follow the same approach used by R09 to quantify
the systematic uncertainty in the determination of $H_0$, by measuring the
impact of a number of variants in the modeling of the Cepheid and SN~Ia data.

In Table 4 we show 15 variants of the previously described analysis for every
combination of choices of distance anchors (NGC\,4258, Milky Way, or LMC), any
two of the preceding or all three; these amount to a total of 105
combinations. Our primary analysis for any anchor choice is given in the first
row (shown in bold) for which that choice initially appears. Column (1) gives
the value of $\chi^2_{\nu}$, Column (2) the number of Cepheids in the fit,
Column (3) the value and total uncertainty in $H_0$, and Column (4) whether the
near-IR data for Cepheids with periods shorter than the completeness limit from
the optical selection were included. Column (5) gives the SN Ia
magnitude-redshift intercept parameter, Column (6) gives the determination of
$M_V^0$ which is specific to the light-curve fitter employed, Column (7) the
calibration system for the metal abundances Column (8) the value and uncertainty
in the metallicity dependence, and Column (9) the value and uncertainty of the
slope of the Cepheid \PL or $P$--$W$ relation. Column (10) gives the minimum
SN~Ia redshift used to define the $m$--$z$ relation, Column (11) encodes aspects
of the SN fitting routine and assumptions therein addressed below, and Column
(12) is the choice of anchors to set the distance scale. Column (13) gives the
type of \PL relation employed, either Wesenheit ($H,V,I$) or $H$-band
only. Column (14) is the reddening law value used for the Cepheids. Column (15)
lists the filters allowed for fitting the SN~Ia light curves and column (16)
gives the value of $R_V$ used to fit the SN light curves.

\subsection{Cepheid Systematics}

In the preceding analysis of the Cepheid data, differences in the determination
of $H_0$ may result from the following variants in the primary analysis: (1)
retention of Cepheids with periods below the optical incompleteness limit; (2)
not allowing for a metallicity dependence; (3) changing the Cepheid reddening
law from $R_V = 3.1$ to $R_V = 2.5$; (4) using only near-IR magnitudes without
reddening corrections; (5) no rejection of outliers in the \PL relations; and
(6) a change in the calibration of chemical abundances.  Each of these changes
was implemented as a variant of the primary analysis with results given in
Table 4. The rationale for the primary analysis over each variant was discussed
in detail in \S 4 of R09, with the exception of (6) which is discussed below.

Taken individually, these variants result in $H_0$ rising or declining by
$\lesssim 1.0$ km s$^{-1}$ Mpc$^{-1}$, which is less than half of the
statistical uncertainty.  A variant resulting in a larger change occurs when we
do not reject Cepheids which are outliers on the \PL relation, raising $H_0$ by
1.3 km s$^{-1}$ Mpc$^{-1}$.  However, the value of $\chi^2_{\nu}$ also triples,
with a total increase in $\chi^2$ of 6 per rejected outlier.  As we expect
outliers {\it a priori} to arise from blending or misidentification of Cepheids
(type or period), resulting in residuals in excess of the typical uncertainty,
we believe it is most sensible to reject them to minimize their impact on the
global solution.  The use of higher or lower thresholds for outlier rejection
has even less impact than including all outliers.  Lowering the outlier
threshold to $2.25\sigma$ (and its accompanying residual magnitude)
reduces $H_0$ by 0.1 km s$^{-1}$ Mpc$^{-1}$.  Raising the threshold to $3.0$
or $4.0\sigma$ reduces $H_0$ by 1.0 or 0.8 km s$^{-1}$ Mpc$^{-1}$,
respectively.  Neglecting a reddening correction for the Cepheids also raises
$H_0$ by 1.3 km s$^{-1}$ Mpc$^{-1}$ but we believe this correction is
warranted.

As an alternative to rejecting outliers we also considered the approach of
simultaneously modeling the distribution of Cepheids and the outliers.
Following \citet{kunz07} we allowed for a nuisance population
of sources along the \PL relation characterized by a broader distribution
($\sigma=1$ mag) and an intercept independent from that of classical Cepheids.
The {\it a posteri} likelihood function for the intercepts of the Cepheid hosts
was then compared to that derived from outlier rejection.  The mean zeropoint of
the SN hosts is greater by 0.013 $\pm$ 0.012 mag.  The mean uncertainty of the
intercepts are a factor of 1.38 greater than those from outlier rejection but
still small compared to the distance precision of each SN.  The only difference
of note (i.e., $> 0.03$ mag) was for the intercept of NGC 4536 which was greater
by $0.08\pm 0.05$ mag in the outlier modeling over the use of rejection.

The chemical abundance values for the Cepheids used in R09 and here were
estimated from nebular lines in \ion{H}{2} regions of the Cepheid hosts using
the R$_{23}$ parameter and the transformation to an oxygen abundance following
\citet[][hereafter ZKH]{zaritsky94}.  There are several alternative
calibrations of the transformation from R$_{23}$ to $\log\,[O/H]$
\citep[e.g.][]{mcgaugh91,pilyugin05}, but these primarily affect the absolute
normalization of the metallicity scale and do not alter the relative
host-to-host differences in abundance or the determination of $H_0$. Recently,
\citet{bresolin11} has redetermined the abundance gradient of NGC 4258 by
adopting the \citet{pilyugin05} calibration of R$_{23}$. This calibration
yields abundances that are consistent with those determined directly by
\citet{bresolin11} in 4 \ion{H}{2} regions in the outer disk of NGC\,4258,
measuring the electron temperature ($T_e$) via the auroral line [\ion{O}{3}]
$\lambda$ 4363. Given this agreement, \citet{bresolin11} suggests the adoption
of a so-called ``T$_e$ scale'' for the determination of absolute chemical
abundances of extragalactic Cepheids.

The T$_e$ recalibration of nebular oxygen abundances not only reduces the
values of $\log\,[O/H]$ by $\sim$ 0.4 dex at the metal-rich end but also
compresses the abundance scale by a factor of 0.69. Based on this scale and
consistent atomic data, \citet{bresolin11} finds a nebular oxygen abundance for
the LMC of $12+\log\,[O/H]=8.36$, moderately lower than the ``canonical'' value
of 8.5 in the ZKH scale. On the T$_e$ scale, the mean apparent metallicity of
the SN Ia and maser hosts would be $12+log\,[O/H]=8.42$; this is closer to the
LMC Cepheids than to the Milky Way Cepheids and a departure from the ZKH scale.
While the abundances of Milky Way Cepheids are not measured the same way (i.e.,
they are based on stellar absorption lines rather than on nearby ionized gas),
they have been directly measured to be $\sim$ 0.3 dex higher than those of LMC
Cepheids \citep{andrievsky02,romaniello08}.  The resulting estimate of 8.66 for
the MW Cepheids on the T$_e$ scale would agree well with recent estimates of
the solar oxygen abundance of 8.69 \citep{asplund09} together with a small
gradient in metallicity away from the solar neighborhood.  This LMC to MW
Cepheid abundance difference of 0.3~dex also agrees well with the T$_e$ scale
compression of the 0.4~dex difference on the ZKH scale for which the value for
MW Cepheids was taken (here and in R09) to be 8.9.

We determined the effect on $H_0$ of a change from the ZKH scale to the T$_e$
scale by transforming the values of 12+log$\,[O/H]$ using equation (3) of
\citet{bresolin11} and assigning values of 8.36 and 8.66 to LMC and MW
Cepheids, respectively.  As seen in Table 4, the value of $H_0$ increases by
0.4 km s$^{-1}$ Mpc$^{-1}$ when using all 3 calibrators and increases by less
than 1.0 km s$^{-1}$ Mpc$^{-1}$ for any combination of 2 calibrators.  The
biggest change, an increase of 2.0 km s$^{-1}$ Mpc$^{-1}$, occurs when only the
MW is used to calibrate the first rung, a direct consequence of the increase of
the metallicity difference between the SN Ia host and MW Cepheids on the T$_e$
scale.  In the presence of uncertainties concerning the appropriate values of
Cepheid abundances, the determination of $H_0$ based on {\it infrared}
observations of Cepheids should be significantly less sensitive to metallicity
differences than optical Cepheid data \citep{marconi05}.  Indeed, the
metallicity correction empirically determined here, $-0.10\pm 0.09$ mag/dex
(using all 3 calibrations) is less than half the value of $\sim -0.25$ mag/dex
measured at optical wavelengths \citep{kennicutt98,sakai04} and its absolute
value is not significant.  A better determination of the difference in
metallicity between MW and extragalactic Cepheids may not occur until the
launch of JWST.

\subsection{ SN Systematics}

Here we consider the following variants in the analysis of the SN~Ia data: (1)
minimum range of SN~Ia $m$--$z$ relation lowered from $z = 0.023$ to $z = 0.01$,
(2) discarding $U$-band SN~Ia light-curve data (fit 61), (3) SN~Ia reddening
parameter $R_V=1.5,2.0,3.1$ (fits 29, 28, and 20), (4) use of a SN~Ia
luminosity-color correction with no prior (i.e., as in the $\beta$ parameter of
SALT II instead of an extinction parameter, $R_V$, in MLCS2k2) (fit 26), (5) a
host-galaxy extinction likelihood prior from galaxy simulations (fit 27), and
(6) use of the SALT-II light-curve fitter (fit 42).  The motivation for these
variants is described in greater detail in R09.

As seen in Table 4, none of these variants taken individually alters the value
of $H_0$ by more than $\sim 1.5$ km s$^{-1}$ Mpc$^{-1}$ from the preferred
solution, less than half the statistical uncertainty.  One of the more
noteworthy variants is the use of the SALT-II light-curve fitter \citep{guy05}
{\it in lieu} of MLCS2k2, since the result of this change can be substantial
for high-redshift data \citep{kessler09}.  Observations of high-redshift SNe~Ia
typically have lower signal-to-noise ratios, and thus place greater reliance on
fitters and on the assumptions they include (e.g., the relation between SN~Ia
color and distance).  In contrast, the determination of $H_0$ is quite {\it
  insensitive} to the fitter; the use of SALT-II results in an increase in
$H_0$ of 1 km s$^{-1}$ Mpc$^{-1}$.

The dispersion of the 15 different determinations of $H_0$ is 0.7 or 0.8 km
s$^{-1}$ Mpc$^{-1}$ for any selected pair of sources of the absolute distance
scale. Adding this measure of analysis systematics to the previous yields
\hofin, a \uncfin\ uncertainty, our best determination.

\begin{table}[h]
\tablenum{5}
\begin{small}
\begin{center}
\vspace{0.4cm}
\begin{tabular}{llllll}
\multicolumn{6}{c}{Table 5: $H_0$ Error Budget for Cepheid and SN~Ia Distance Ladders$^*$} \\
\hline
\hline
Term & Description &  Previous & R09  & Here & Here \\
%\multicolumn{3}{c}{ } & \multicolumn{3}{l}{anchor}\\
          &                    & LMC &  N4258 &   N4258 & All 3 \\
\hline
$\sigma_{\rm anchor}$  &   Anchor distance  &  5\%  & 3\% & 3\% & 1.3\% \\
$\sigma_{{\rm anchor}-PL}$  &  Mean of \PL in anchor & 2.5\%  &  1.5\% & 1.4\% & 0.7\%$^a$ \\
$\sigma_{{\rm host}-PL}/\sqrt{n}$  &  Mean of \PL values in SN hosts  & 1.5\% & 1.5\% & 0.6 \% & 0.6\% \\
$\sigma_{\rm SN}/\sqrt{n}$  &  Mean of SN~Ia calibrators &  2.5\% & 2.5\% & 1.9\% & 1.9\% \\
$\sigma_{m-z}$  &  SN~Ia $m$--$z$ relation & 1\% & 0.5\%  & 0.5\% & 0.5\% \\
$R \sigma_{\lambda,1,2}$  & Cepheid reddening, zeropoints, anchor-to-hosts & 4.5\% & 0.3\% & 0.0\% & 1.4\% \\
$\sigma_{Z}$ & Cepheid metallicity, anchor-to-hosts  & 3\% & 1.1\% & 0.6 \%  & 1.0\%  \\
$\sigma_{\rm PL}$ & \PL slope, $\Delta$ log $P$, anchor-to-hosts & 4\% & 0.5\%  &  0.4\% & 0.6\% \\
$\sigma_{\rm WFPC2}$ & WFPC2 CTE, long-short & 3\% & 0\% & 0\%  & 0\% \\
\hline
\multicolumn{2}{l}{subtotal, $\sigma_{H_0}$} &     10\%  &  4.7 \% & 4.0\%  & 2.9\% \\
\hline
\multicolumn{2}{l}{Analysis Systematics} & NA & 1.3\% & 1.0\% & 1.0\% \\
 \hline
\multicolumn{2}{l}{Total, $\sigma_{H_0}$} &     10\%  &  4.8 \% & 4.1\%  & 3.1\% \\
\hline
 \hline
 \multicolumn{6}{l}{$^*$Derived from diagonal elements of the
   covariance matrix propagated via the error matrices associated} \\
\multicolumn{6}{l}{\ \ with Equations 1, 3, 7, and 8.} \\
 \multicolumn{6}{l}{$^a$For Milky Way parallax, this term is already
   included with the term above.} \\
\end{tabular}
\end{center}
\end{small} 
\end{table}

\section{Dark Energy and Neutrinos}

An independent and precise measurement of $H_0$ is an important complement to
the determination of cosmological model parameters. Alternatively, it serves as
a powerful test of model-constrained measurements at higher redshifts.  It is
beyond the scope of this paper to provide a complete analysis of the impact of
the measurement of $H_0$ on the cosmological model from all extant data.  We
encourage others to do so.  However, one such example using the present
measurement of $H_0$ can be illustrative.

Making use of the simplest present hypothesis for the cosmological model
(namely $\Lambda$-cold-dark-matter without curvature, exotic neutrino physics,
or specific early-Universe physics), and using the single most powerful
cosmological data set (the 7-year WMAP results from \citealt{komatsu11}),
results in a predicted value of $H_0 = 71.0 \pm 2.5$ km s$^{-1}$ Mpc$^{-1}$.
This value agrees well with our determination of \hofin  at better than the
combined 1$\sigma$ confidence level.

Alternatively, we can use the WMAP data together with the measured value of
$H_0$ to constrain added complexity to the model.  In Figure 10 we show the use
of this data combination for constraining a redshift-independent dark energy
equation-of-state parameter ($w$), the number of relativistic species (e.g.,
neutrino number), and the sum of neutrino masses.  The result for dark energy
is $w = -1.08 \pm 0.10$, about 20\% more precise than the same result derived
from the determination of $H_0$ in R09.  If we had perfect knowledge of the
CMB, our overall 30\% increase in the precision of $H_0$ would yield the
same-sized improvement in the determination of $w$. However, the fractional
uncertainty in $\Omega_M h^2$ from the WMAP 7-year analysis is comparable to
our measurement of $H_0$; thus, greater precision in $w$ may still be wrung
from future higher-precision measurements of the CMB by WMAP or {\it Planck}.

The enhanced precision in measuring $H_0$ also provides a strong rebuff to
recent attempts to explain accelerated expansion without dark energy but
rather by our presence in the center of a massive void of gigaparsec scale.
Already such models are hard to fathom as they require an exotic location for
the observer, at the center of the void to within a part in a million
\citep{blomqvist10} to avoid an excess dipole in the CMB.  It is also not yet
apparent if such a model is consistent with other observables of the CMB or the
late-time integrated Sachs-Wolfe effect.  However, using measurements of
$H(z>1)$ to constrain void models of the Lemaitre, Tolmon, and Bondi variety
already predicts slower-than-observed local expansion with values of $H_0$=60
\citep{nadathur10} or 62 \citep{wiltshire07} km s$^{-1}$ Mpc$^{-1}$, more than
5 $\sigma$ below our measurement.

Comparable improvements to cosmological constraints on relativistic species are
also realized from R09, as shown in Figure 10.  Most interesting may be the
effective number of relativistic species, $N_{\rm eff} = 4.2\pm0.75$, which is
nominally higher than the value of 3.046 expected from the three known neutrino
species plus tau-neutrino heating from $e^+ e^-$ collisions \citep{mangano05}.
While this nominal excess of relativistic species has been noted previously
\citep[e.g.,]{reid10,komatsu11,dunkley10}, and even interpreted as a possible
indication of the presence of a sterile neutrino \citep{hamann10}, we caution
that the cosmological model provides other avenues for reducing the
significance of this result including additional degrees of freedom for
curvature, dark energy, primordial helium abundance, and neutrino masses.  The
30\% improvement in the present constraint on $H_0$ combined with improved high
resolution CMB data (e.g., \cite{dunkley10}) and ultimately with {\it Planck}
satellite CMB data should reduce the present uncertainty in $N_{\rm eff} $ by a
factor of $\sim$ 3 which may provide a more definitive conclusion on the
presence of excess radiation in the early Universe.
  
\section{Discussion}

Examination of the complete error budget for $H_0$ in the last two columns of
Table 5 indicates additional approaches for improved precision in future
measurements of $H_0$.  Expanding the sample of well-measured parallaxes to
Milky Way Cepheids (especially those at log\,$P > 1$) with the GAIA satellite
could drive the precision of the first rung of the distance ladder well under
1\%.  However, as we have found with the ``baker's dozen'' of present Milky Way
parallaxes, much of this precision would be lost without better
cross-calibration between the space and ground photometric systems used to
measure Cepheids, near and far.

The largest remaining term comes from the quite limited sample of ideal SN~Ia
calibrators, just 8 objects.  The occurrence of an ideal SN~Ia in the small
volume within which {\it HST} can measure Cepheids ($R \approx 30$~Mpc) is
rare, on average only once every 2--3 yr.  Given the recent proliferation of SN
surveys and instances of multiple, independent discoveries, we are confident
that all such SNe~Ia within this volume are being found.  Collecting more will
require extending the range of Cepheid measurements --- without introducing new
systematics --- and patience.  The forthcoming {\it James Webb Space Telescope
  (JWST)} offers a promising route to extend Cepheid observations out to 50~Mpc
and to redder wavelengths, where uncertainties due to possible variations in
the extinction law and the dependence of Cepheid luminosities on metallicity
are further reduced.  This extension would increase the SN sample suitable for
calibration by a factor of $\sim 5$, reaching $\sim 40$ ideal SNe~Ia observed
over the past 20 yr.  Based on a 5\% distance precision per ideal SN, such a
sample would enable a determination of $H_0$ to better than 1\%.  However,
discovering these Cepheids may require imaging at optical wavelengths where the
amplitude of the variations is significant, a requirement which will challenge
the short-wavelength capabilities of {\it JWST}.

\section{Summary and Conclusions}

We have improved upon the precision of the measurement of $H_0$ from
\citet{riess09a} by (1) more than doubling the sample of Cepheids observed in
the near-IR in SN~Ia host galaxies, (2) expanding the SN~Ia sample from 6 to 8
with the addition of SN 2007af and SN 2007sr, (3) increasing the sample of
Cepheids observed in NGC\,4258 by 20\%, (4) reducing the difference in
metallicity for the observed sample of Cepheids between the calibrator and the
SN hosts, and (5) calibrating all optical Cepheid colors with WFC3 to remove
cross-instrument zeropoint errors.  Further improvements to the precision and
reliability of the measurement of $H_0$ come from the use of additional sources
of calibration for the first rung, foremost of these are the trigonometric
parallaxex of 13 Cepheids in the Milky Way.

Our primary analysis gives $H_0 =$ \hofin\ including systematic errors
determined from varying assumptions and priors used in the analysis. The
combination of this result alone with the WMAP 7-year constraints yields $w =
-1.08 \pm 0.10$ and improves constraints on a possible but still uncertain
excess in relativistic species above the number of known neutrino
flavors. The measured $H_0$ is also highly inconsistent with the simplest inhomogeneous matter models
invoked to explain the apparent acceleration of the Universe without dark energy.
Given that statistical errors still dominate over systematic errors,
future work is likely to further improve the precision of the determination of
$H_0$.

\bigskip 
\medskip 

We are grateful to William Januszewski for his help in executing this program
on {\it HST}.  We are indebted to Mike Hudson for assisting with the
peculiar-velocity calculations from the PSCz survey, to David Larson for
contributions to the WMAP MCMC analysis, to Daniel Scolnic for donating some
useful routines, and to Mark Huber for an analysis of pre-discovery
observations of SN 2007sr.  We thank Chris Kochanek and Kris Stanek
for their support of GO-11570.  Financial support for this work was provided by
NASA through programs GO-11570 and GO-10802 from the Space Telescope Science
Institute, which is operated by AURA, Inc., under NASA contract NAS
5-26555. A.V.F.'s supernova group at U.~C. Berkeley is also supported by NSF
grant AST--0607485 and by the TABASGO Foundation.  L.M.M. acknowledges support
from a Texas A\&M University faculty startup fund.  The metallicity
measurements for NGC 5584 and NGC 4038/9 presented herein were obtained with
the W.~M. Keck Observatory, which is operated as a scientific partnership among
the California Institute of Technology, the University of California, and NASA;
the observatory was made possible by the generous financial support of the
W.~M. Keck Foundation.

\vfill
\eject

{\bf Figure Captions}

Figure 1: {\it HST} observations of the host galaxies used to measure
$H_0$.  The data employed to observe Cepheids in 8 SN~Ia hosts and
NGC\,4258 have been collected over 15~yr with 4 cameras over $\sim
500$ orbits of {\it HST} time.  Two-month long campaigns in $F555W$
and $F814W$ were initially used to discover Cepheids from their light
curves.  Subsequent follow-up observations in $F555W$ enabled the
discovery of Cepheids with $P > 60$~d. Near-IR follow-up data have
been used to reduce the effects of host-galaxy extinction and
sensitivity to metallicity.

Figure 2: {\it HST} images of NGC\,5584.  The positions of Cepheids
with periods in the range $P > 60$~d, $30 < P < 60$~d, and $10 < P <
30$~d are indicated by red, blue, and green circles, respectively.  A
yellow circle indicates the position of the host galaxy's SN~Ia.  
The orientation is indicated by the compass rose whose vectors have
lengths of 15$\arcsec$.  The black and white regions of the images
show the WFC3 optical data and the color includes the WFC3-IR data.

Figure 3.  As in Figure 2, for NGC\,4038/4039.

Figure 4: {\it HST} WFC3-$F160W$ image of NGC\,3370.  Upper panel: The
positions of Cepheids with periods in the range $P > 60$~d, $30 < P < 60$~d,
and $10 < P < 30$~d are indicated by red, blue, and green circles,
respectively.  A yellow circle indicates the position of the host galaxy's
SN~Ia.  The orientation is indicated by the compass rose whose vectors have
lengths of 15$\arcsec$.  The fields of view for the NIC2 follow-up fields from
\citet{riess09a} are indicated.  Lower Panel: close-up showing the field of
NGC3370-blue as observed with WFC3-IR (left) and with 4.7 times more exposure
time with NIC2 (right).

Figure 5: Example of scene modeling for the $\sim 1''$ surrounding
typical short, medium, and long-period Cepheids in one WFC3 field,
NGC\,5584.  For each Cepheid, the stamp on the left shows the region
around the Cepheid, the middle stamp shows the model of the stellar
sources, and the right stamp is the residual of the image minus the
model.  The position of the Cepheid as determined from the optical
data is indicated by the circle.

Figure 6: WFC3-IR versus NIC2 $F160W$ Cepheid photometry.  Some of the apparent
dispersion results from the random phases of the Cepheids observed with WFC3.

Figure 7: Near-IR Cepheid period-luminosity relations.  For the 8
SN~Ia hosts and the distance-scale anchor, NGC\,4258, the Cepheid
magnitudes are from the same instrument and filter combination, WFC3
$F160W$.  This uniformity allows for a significant reduction in
systematic error when utilizing the difference in these relations
along the distance ladder.  The measured metallicity for all of the
Cepheids is comparable to solar (log\,[O/H] $\approx 8.9$).  A single
slope has been fit to the relations and is shown as the solid line.
20\% of the objects were outliers from the relations (open diamonds)
and are flagged as such for the subsequent analysis.  Filled points
with asterisks indicate Cepheids whose periods are shorter than the
incompleteness limit identified from their optical detection.

Figure 8: Relative distances from Cepheids and SNe~Ia.  The bottom
abscissa shows the peak apparent visual magnitude of each SN~Ia (red
points) corrected for reddening and to the fiducial brightness (using
the luminosity vs. light-curve shape relations), $m_V^0$.  The top
abscissa includes the intercept of the $m_V^0$--log\,$cz$ relation for
SNe~Ia, $a_v$ to provide SN~Ia distance measures, $m_V^0 + 5a_v$,
quantities which are independent of the choice of a fiducial SN Ia.  The
right-hand ordinate shows the relative distances between the hosts
determined from the Cepheid $VIH$ Wesenheit relations.  The left
ordinate shows the same thing, with the addition of the independent
geometric distance to NGC\,4258 (blue point) based on its
circumnuclear megamasers.  The contribution of the nearby SN~Ia and
Cepheid data to $H_0$ can be expressed as a determination of
$m_{V,4258}^0$, the theoretical mean of 8 fiducial SNe~Ia in
NGC\,4258.

Figure 9: Uncertainties in the determination of the Hubble constant.
Uncertainties are squared to show their contribution to the quadrature sum.
These terms are given in Table 5.

Figure 10: Confidence regions in the plane of $H_0$ and the equation-of-state
parameter of dark energy, $w$ and neutrino properties.  The localization of the
third acoustic peak in the WMAP 7-year data \citep{komatsu11} produces a
confidence region which is narrow but highly degenerate with the dark energy
equation of state (upper panel).  The improved measurement of $H_0$, \hofin,
from the SH0ES program is complementary to the WMAP constraint, resulting in a
determination of $w = -1.08 \pm 0.10$ assuming a constant $w$.  This result is
comparable in precision to determinations of $w$ from baryon acoustic
oscillations and high-redshift SNe~Ia, but is independent of both.  The inner
regions are 68\% confidence and the outer regions are 95\% confidence.  The
modest tilt of the SH0ES measurement of 0.2\% in $H_0$ for a change in $w$ of
0.1 shown as the dotted lines in the upper panel, results from the mild
dependence of $a_v$ on $w$, corresponding to the change in $H$ for changes from
$w=-1$ at the mean SN rdshift of $z = 0.04$.  The measurement of $H_0$ is made
at $j_0 = 1$ (i.e., $w = -1$).  Constraints on the mass and number of
relativisitic species (e.g., neutrinos) are shown in the middle and lower
panels, respectively.

\bibliographystyle{apj}
\bibliography{ariess}

\begin{thebibliography}{81}
\expandafter\ifx\csname natexlab\endcsname\relax\def\natexlab#1{#1}\fi

\bibitem[{{Andrievsky} {et~al.}(2002){Andrievsky}, {Kovtyukh}, {Luck},
  {L{\'e}pine}, {Bersier}, {Maciel}, {Barbuy}, {Klochkova}, {Panchuk}, \&
  {Karpischek}}]{andrievsky02}
{Andrievsky}, S.~M., {Kovtyukh}, V.~V., {Luck}, R.~E., {L{\'e}pine}, J.~R.~D.,
  {Bersier}, D., {Maciel}, W.~J., {Barbuy}, B., {Klochkova}, V.~G., {Panchuk},
  V.~E., \& {Karpischek}, R.~U. 2002, \aap, 381, 32

\bibitem[{{Argon} {et~al.}(2007){Argon}, {Greenhill}, {Reid}, {Moran}, \&
  {Humphreys}}]{argon07}
{Argon}, A.~L., {Greenhill}, L.~J., {Reid}, M.~J., {Moran}, J.~M., \&
  {Humphreys}, E.~M.~L. 2007, \apj, 659, 1040

\bibitem[{{Asplund} {et~al.}(2009){Asplund}, {Grevesse}, {Sauval}, \&
  {Scott}}]{asplund09}
{Asplund}, M., {Grevesse}, N., {Sauval}, A.~J., \& {Scott}, P. 2009, \araa, 47,
  481

\bibitem[{{Benedict} {et~al.}(2007){Benedict}, {McArthur}, {Feast}, {Barnes},
  {Harrison}, {Patterson}, {Menzies}, {Bean}, \& {Freedman}}]{benedict07}
{Benedict}, G.~F., {McArthur}, B.~E., {Feast}, M.~W., {Barnes}, T.~G.,
  {Harrison}, T.~E., {Patterson}, R.~J., {Menzies}, J.~W., {Bean}, J.~L., \&
  {Freedman}, W.~L. 2007, \aj, 133, 1810

\bibitem[{{Blomqvist} \& {M{\"o}rtsell}(2010)}]{blomqvist10}
{Blomqvist}, M. \& {M{\"o}rtsell}, E. 2010, \jcap, 5, 6

\bibitem[{{Branchini} {et~al.}(1999){Branchini}, {Teodoro}, {Frenk},
  {Schmoldt}, {Efstathiou}, {White}, {Saunders}, {Sutherland},
  {Rowan-Robinson}, {Keeble}, {Tadros}, {Maddox}, \& {Oliver}}]{branchini99}
{Branchini}, E., {Teodoro}, L., {Frenk}, C.~S., {Schmoldt}, I., {Efstathiou},
  G., {White}, S.~D.~M., {Saunders}, W., {Sutherland}, W., {Rowan-Robinson},
  M., {Keeble}, O., {Tadros}, H., {Maddox}, S., \& {Oliver}, S. 1999, \mnras,
  308, 1

\bibitem[{{Bresolin}(2011)}]{bresolin11}
{Bresolin}, F. 2011, \apj, {in press}

\bibitem[{{Cardelli} {et~al.}(1989){Cardelli}, {Clayton}, \&
  {Mathis}}]{cardelli89}
{Cardelli}, J.~A., {Clayton}, G.~C., \& {Mathis}, J.~S. 1989, \apj, 345, 245

\bibitem[{{Dunkley} {et~al.}(2010){Dunkley}, {Hlozek}, {Sievers}, {Acquaviva},
  {Ade}, {Aguirre}, {Amiri}, {Appel}, {Barrientos}, {Battistelli}, {Bond},
  {Brown}, {Burger}, {Chervenak}, {Das}, {Devlin}, {Dicker}, {Bertrand
  Doriese}, {Dunner}, {Essinger-Hileman}, {Fisher}, {Fowler}, {Hajian},
  {Halpern}, {Hasselfield}, {Hernandez-Monteagudo}, {Hilton}, {Hilton},
  {Hincks}, {Huffenberger}, {Hughes}, {Hughes}, {Infante}, {Irwin}, {Juin},
  {Kaul}, {Klein}, {Kosowsky}, {Lau}, {Limon}, {Lin}, {Lupton}, {Marriage},
  {Marsden}, {Mauskopf}, {Menanteau}, {Moodley}, {Moseley}, {Netterfield},
  {Niemack}, {Nolta}, {Page}, {Parker}, {Partridge}, {Reid}, {Sehgal},
  {Sherwin}, {Spergel}, {Staggs}, {Swetz}, {Switzer}, {Thornton}, {Trac},
  {Tucker}, {Warne}, {Wollack}, \& {Zhao}}]{dunkley10}
{Dunkley}, J., {Hlozek}, R., {Sievers}, J., {Acquaviva}, V., {Ade}, P.~A.~R.,
  {Aguirre}, P., {Amiri}, M., {Appel}, J.~W., {Barrientos}, L.~F.,
  {Battistelli}, E.~S., {Bond}, J.~R., {Brown}, B., {Burger}, B., {Chervenak},
  J., {Das}, S., {Devlin}, M.~J., {Dicker}, S.~R., {Bertrand Doriese}, W.,
  {Dunner}, R., {Essinger-Hileman}, T., {Fisher}, R.~P., {Fowler}, J.~W.,
  {Hajian}, A., {Halpern}, M., {Hasselfield}, M., {Hernandez-Monteagudo}, C.,
  {Hilton}, G.~C., {Hilton}, M., {Hincks}, A.~D., {Huffenberger}, K.~M.,
  {Hughes}, D.~H., {Hughes}, J.~P., {Infante}, L., {Irwin}, K.~D., {Juin},
  J.~B., {Kaul}, M., {Klein}, J., {Kosowsky}, A., {Lau}, J.~M., {Limon}, M.,
  {Lin}, Y., {Lupton}, R.~H., {Marriage}, T.~A., {Marsden}, D., {Mauskopf}, P.,
  {Menanteau}, F., {Moodley}, K., {Moseley}, H., {Netterfield}, C.~B.,
  {Niemack}, M.~D., {Nolta}, M.~R., {Page}, L.~A., {Parker}, L., {Partridge},
  B., {Reid}, B., {Sehgal}, N., {Sherwin}, B., {Spergel}, D.~N., {Staggs},
  S.~T., {Swetz}, D.~S., {Switzer}, E.~R., {Thornton}, R., {Trac}, H.,
  {Tucker}, C., {Warne}, R., {Wollack}, E., \& {Zhao}, Y. 2010, ArXiv e-prints

\bibitem[{{Fitzpatrick} {et~al.}(2002){Fitzpatrick}, {Ribas}, {Guinan},
  {DeWarf}, {Maloney}, \& {Massa}}]{fitzpatrick02}
{Fitzpatrick}, E.~L., {Ribas}, I., {Guinan}, E.~F., {DeWarf}, L.~E., {Maloney},
  F.~P., \& {Massa}, D. 2002, \apj, 564, 260

\bibitem[{{Fitzpatrick} {et~al.}(2003){Fitzpatrick}, {Ribas}, {Guinan},
  {Maloney}, \& {Claret}}]{fitzpatrick03}
{Fitzpatrick}, E.~L., {Ribas}, I., {Guinan}, E.~F., {Maloney}, F.~P., \&
  {Claret}, A. 2003, \apj, 587, 685

\bibitem[{{Fouqu{\'e}} {et~al.}(2007){Fouqu{\'e}}, {Arriagada}, {Storm},
  {Barnes}, {Nardetto}, {M{\'e}rand}, {Kervella}, {Gieren}, {Bersier},
  {Benedict}, \& {McArthur}}]{fouque07}
{Fouqu{\'e}}, P., {Arriagada}, P., {Storm}, J., {Barnes}, T.~G., {Nardetto},
  N., {M{\'e}rand}, A., {Kervella}, P., {Gieren}, W., {Bersier}, D.,
  {Benedict}, G.~F., \& {McArthur}, B.~E. 2007, \aap, 476, 73

\bibitem[{{Freedman} \& {Madore}(2010)}]{freedman10}
{Freedman}, W.~L. \& {Madore}, B.~F. 2010, \araa, 48, 673

\bibitem[{{Freedman} {et~al.}(2001){Freedman}, {Madore}, {Gibson}, {Ferrarese},
  {Kelson}, {Sakai}, {Mould}, {Kennicutt}, {Ford}, {Graham}, {Huchra},
  {Hughes}, {Illingworth}, {Macri}, \& {Stetson}}]{freedman01}
{Freedman}, W.~L., {Madore}, B.~F., {Gibson}, B.~K., {Ferrarese}, L., {Kelson},
  D.~D., {Sakai}, S., {Mould}, J.~R., {Kennicutt}, Jr., R.~C., {Ford}, H.~C.,
  {Graham}, J.~A., {Huchra}, J.~P., {Hughes}, S.~M.~G., {Illingworth}, G.~D.,
  {Macri}, L.~M., \& {Stetson}, P.~B. 2001, \apj, 553, 47

\bibitem[{{Ganeshalingam} {et~al.}(2010){Ganeshalingam}, {Li}, {Filippenko},
  {Anderson}, {Foster}, {Gates}, {Griffith}, {Grigsby}, {Joubert}, {Leja},
  {Lowe}, {Macomber}, {Pritchard}, {Thrasher}, \& {Winslow}}]{ganeshalingam10}
{Ganeshalingam}, M., {Li}, W., {Filippenko}, A.~V., {Anderson}, C., {Foster},
  G., {Gates}, E.~L., {Griffith}, C.~V., {Grigsby}, B.~J., {Joubert}, N.,
  {Leja}, J., {Lowe}, T.~B., {Macomber}, B., {Pritchard}, T., {Thrasher}, P.,
  \& {Winslow}, D. 2010, \apjs, 190, 418

\bibitem[{{Gibson}(2000)}]{gibson00a}
{Gibson}, B.~K. 2000, \memsai, 71, 693

\bibitem[{{Gibson} {et~al.}(2000){Gibson}, {Stetson}, {Freedman}, {Mould},
  {Kennicutt}, {Huchra}, {Sakai}, {Graham}, {Fassett}, {Kelson}, {Ferrarese},
  {Hughes}, {Illingworth}, {Macri}, {Madore}, {Sebo}, \&
  {Silbermann}}]{gibson00b}
{Gibson}, B.~K., {Stetson}, P.~B., {Freedman}, W.~L., {Mould}, J.~R.,
  {Kennicutt}, Jr., R.~C., {Huchra}, J.~P., {Sakai}, S., {Graham}, J.~A.,
  {Fassett}, C.~I., {Kelson}, D.~D., {Ferrarese}, L., {Hughes}, S.~M.~G.,
  {Illingworth}, G.~D., {Macri}, L.~M., {Madore}, B.~F., {Sebo}, K.~M., \&
  {Silbermann}, N.~A. 2000, \apj, 529, 723

\bibitem[{{Greenhill}(2009)}]{greenhill09}
{Greenhill}, L. 2009, in ArXiv Astrophysics e-prints, Vol. 2010, astro2010: The
  Astronomy and Astrophysics Decadal Survey, 103

\bibitem[{{Groenewegen}(1999)}]{groenewegen99}
{Groenewegen}, M.~A.~T. 1999, \aaps, 139, 245

\bibitem[{{Guinan} {et~al.}(1998){Guinan}, {Fitzpatrick}, {Dewarf}, {Maloney},
  {Maurone}, {Ribas}, {Pritchard}, {Bradstreet}, \& {Gim{\'e}nez}}]{guinan98}
{Guinan}, E.~F., {Fitzpatrick}, E.~L., {Dewarf}, L.~E., {Maloney}, F.~P.,
  {Maurone}, P.~A., {Ribas}, I., {Pritchard}, J.~D., {Bradstreet}, D.~H., \&
  {Gim{\'e}nez}, A. 1998, \apjl, 509, L21

\bibitem[{{Guy} {et~al.}(2005){Guy}, {Astier}, {Nobili}, {Regnault}, \&
  {Pain}}]{guy05}
{Guy}, J., {Astier}, P., {Nobili}, S., {Regnault}, N., \& {Pain}, R. 2005,
  \aap, 443, 781

\bibitem[{{Hamann} {et~al.}(2010){Hamann}, {Hannestad}, {Raffelt}, {Tamborra},
  \& {Wong}}]{hamann10}
{Hamann}, J., {Hannestad}, S., {Raffelt}, G.~G., {Tamborra}, I., \& {Wong},
  Y.~Y.~Y. 2010, Physical Review Letters, 105, 181301

\bibitem[{{Herrnstein} {et~al.}(1999){Herrnstein}, {Moran}, {Greenhill},
  {Diamond}, {Inoue}, {Nakai}, {Miyoshi}, {Henkel}, \& {Riess}}]{herrnstein99}
{Herrnstein}, J.~R., {Moran}, J.~M., {Greenhill}, L.~J., {Diamond}, P.~J.,
  {Inoue}, M., {Nakai}, N., {Miyoshi}, M., {Henkel}, C., \& {Riess}, A. 1999,
  \nat, 400, 539

\bibitem[{{Hicken} {et~al.}(2009{\natexlab{a}}){Hicken}, {Challis}, {Jha},
  {Kirshner}, {Matheson}, {Modjaz}, {Rest}, {Wood-Vasey}, {Bakos}, {Barton},
  {Berlind}, {Bragg}, {Brice{\~n}o}, {Brown}, {Caldwell}, {Calkins}, {Cho},
  {Ciupik}, {Contreras}, {Dendy}, {Dosaj}, {Durham}, {Eriksen}, {Esquerdo},
  {Everett}, {Falco}, {Fernandez}, {Gaba}, {Garnavich}, {Graves}, {Green},
  {Groner}, {Hergenrother}, {Holman}, {Hradecky}, {Huchra}, {Hutchison},
  {Jerius}, {Jordan}, {Kilgard}, {Krauss}, {Luhman}, {Macri}, {Marrone},
  {McDowell}, {McIntosh}, {McNamara}, {Megeath}, {Mochejska}, {Munoz},
  {Muzerolle}, {Naranjo}, {Narayan}, {Pahre}, {Peters}, {Peterson}, {Rines},
  {Ripman}, {Roussanova}, {Schild}, {Sicilia-Aguilar}, {Sokoloski}, {Smalley},
  {Smith}, {Spahr}, {Stanek}, {Barmby}, {Blondin}, {Stubbs}, {Szentgyorgyi},
  {Torres}, {Vaz}, {Vikhlinin}, {Wang}, {Westover}, {Woods}, \&
  {Zhao}}]{hicken09a}
{Hicken}, M., {Challis}, P., {Jha}, S., {Kirshner}, R.~P., {Matheson}, T.,
  {Modjaz}, M., {Rest}, A., {Wood-Vasey}, W.~M., {Bakos}, G., {Barton}, E.~J.,
  {Berlind}, P., {Bragg}, A., {Brice{\~n}o}, C., {Brown}, W.~R., {Caldwell},
  N., {Calkins}, M., {Cho}, R., {Ciupik}, L., {Contreras}, M., {Dendy}, K.,
  {Dosaj}, A., {Durham}, N., {Eriksen}, K., {Esquerdo}, G., {Everett}, M.,
  {Falco}, E., {Fernandez}, J., {Gaba}, A., {Garnavich}, P., {Graves}, G.,
  {Green}, P., {Groner}, T., {Hergenrother}, C., {Holman}, M.~J., {Hradecky},
  V., {Huchra}, J., {Hutchison}, B., {Jerius}, D., {Jordan}, A., {Kilgard}, R.,
  {Krauss}, M., {Luhman}, K., {Macri}, L., {Marrone}, D., {McDowell}, J.,
  {McIntosh}, D., {McNamara}, B., {Megeath}, T., {Mochejska}, B., {Munoz}, D.,
  {Muzerolle}, J., {Naranjo}, O., {Narayan}, G., {Pahre}, M., {Peters}, W.,
  {Peterson}, D., {Rines}, K., {Ripman}, B., {Roussanova}, A., {Schild}, R.,
  {Sicilia-Aguilar}, A., {Sokoloski}, J., {Smalley}, K., {Smith}, A., {Spahr},
  T., {Stanek}, K.~Z., {Barmby}, P., {Blondin}, S., {Stubbs}, C.~W.,
  {Szentgyorgyi}, A., {Torres}, M.~A.~P., {Vaz}, A., {Vikhlinin}, A., {Wang},
  Z., {Westover}, M., {Woods}, D., \& {Zhao}, P. 2009{\natexlab{a}}, \apj, 700,
  331

\bibitem[{{Hicken} {et~al.}(2009{\natexlab{b}}){Hicken}, {Wood-Vasey},
  {Blondin}, {Challis}, {Jha}, {Kelly}, {Rest}, \& {Kirshner}}]{hicken09b}
{Hicken}, M., {Wood-Vasey}, W.~M., {Blondin}, S., {Challis}, P., {Jha}, S.,
  {Kelly}, P.~L., {Rest}, A., \& {Kirshner}, R.~P. 2009{\natexlab{b}}, \apj,
  700, 1097

\bibitem[{{Hui} \& {Greene}(2006)}]{hui06}
{Hui}, L. \& {Greene}, P.~B. 2006, \prd, 73, 123526

\bibitem[{{Humphreys} {et~al.}(2005){Humphreys}, {Argon}, {Greenhill}, {Moran},
  \& {Reid}}]{humphreys05}
{Humphreys}, E.~M.~L., {Argon}, A.~L., {Greenhill}, L.~J., {Moran}, J.~M., \&
  {Reid}, M.~J. 2005, in Astronomical Society of the Pacific Conference Series,
  Vol. 340, Future Directions in High Resolution Astronomy, ed. {J.~Romney \&
  M.~Reid}, 466

\bibitem[{{Humphreys} {et~al.}(2008){Humphreys}, {Reid}, {Greenhill}, {Moran},
  \& {Argon}}]{humphreys08}
{Humphreys}, E.~M.~L., {Reid}, M.~J., {Greenhill}, L.~J., {Moran}, J.~M., \&
  {Argon}, A.~L. 2008, \apj, 672, 800

\bibitem[{{Jha} {et~al.}(2007){Jha}, {Riess}, \& {Kirshner}}]{jha07}
{Jha}, S., {Riess}, A.~G., \& {Kirshner}, R.~P. 2007, \apj, 659, 122

\bibitem[{{Kalirai} {et~al.}(2009){Kalirai}, {Deustua}, {Baggett}, {Bohlin},
  {Brown}, {MacKenty}, {McCullough}, {Rajan}, {Riess}, {Sabbi}, \&
  {Sirianni}}]{kalirai09}
{Kalirai}, J.~S., {Deustua}, S., {Baggett}, S., {Bohlin}, R., {Brown}, T.,
  {MacKenty}, J., {McCullough}, P., {Rajan}, A., {Riess}, A., {Sabbi}, E., \&
  {Sirianni}, M. 2009, {The Photometric Calibration of WFC3: SMOV and Cycle 17
  Observing Plan}, Tech. rep., Space Telescope Science Institute

\bibitem[{{Kelly} {et~al.}(2010){Kelly}, {Hicken}, {Burke}, {Mandel}, \&
  {Kirshner}}]{kelly10}
{Kelly}, P.~L., {Hicken}, M., {Burke}, D.~L., {Mandel}, K.~S., \& {Kirshner},
  R.~P. 2010, \apj, 715, 743

\bibitem[{{Kennicutt} {et~al.}(1998){Kennicutt}, {Stetson}, {Saha}, {Kelson},
  {Rawson}, {Sakai}, {Madore}, {Mould}, {Freedman}, {Bresolin}, {Ferrarese},
  {Ford}, {Gibson}, {Graham}, {Han}, {Harding}, {Hoessel}, {Huchra}, {Hughes},
  {Illingworth}, {Macri}, {Phelps}, {Silbermann}, {Turner}, \&
  {Wood}}]{kennicutt98}
{Kennicutt}, Jr., R.~C., {Stetson}, P.~B., {Saha}, A., {Kelson}, D., {Rawson},
  D.~M., {Sakai}, S., {Madore}, B.~F., {Mould}, J.~R., {Freedman}, W.~L.,
  {Bresolin}, F., {Ferrarese}, L., {Ford}, H., {Gibson}, B.~K., {Graham},
  J.~A., {Han}, M., {Harding}, P., {Hoessel}, J.~G., {Huchra}, J.~P., {Hughes},
  S.~M.~G., {Illingworth}, G.~D., {Macri}, L.~M., {Phelps}, R.~L.,
  {Silbermann}, N.~A., {Turner}, A.~M., \& {Wood}, P.~R. 1998, \apj, 498, 181

\bibitem[{{Kessler} {et~al.}(2009){Kessler}, {Becker}, {Cinabro}, {Vanderplas},
  {Frieman}, {Marriner}, {Davis}, {Dilday}, {Holtzman}, {Jha}, {Lampeitl},
  {Sako}, {Smith}, {Zheng}, {Nichol}, {Bassett}, {Bender}, {Depoy}, {Doi},
  {Elson}, {Filippenko}, {Foley}, {Garnavich}, {Hopp}, {Ihara}, {Ketzeback},
  {Kollatschny}, {Konishi}, {Marshall}, {McMillan}, {Miknaitis}, {Morokuma},
  {M{\"o}rtsell}, {Pan}, {Prieto}, {Richmond}, {Riess}, {Romani}, {Schneider},
  {Sollerman}, {Takanashi}, {Tokita}, {van der Heyden}, {Wheeler}, {Yasuda}, \&
  {York}}]{kessler09}
{Kessler}, R., {Becker}, A.~C., {Cinabro}, D., {Vanderplas}, J., {Frieman},
  J.~A., {Marriner}, J., {Davis}, T.~M., {Dilday}, B., {Holtzman}, J., {Jha},
  S.~W., {Lampeitl}, H., {Sako}, M., {Smith}, M., {Zheng}, C., {Nichol}, R.~C.,
  {Bassett}, B., {Bender}, R., {Depoy}, D.~L., {Doi}, M., {Elson}, E.,
  {Filippenko}, A.~V., {Foley}, R.~J., {Garnavich}, P.~M., {Hopp}, U., {Ihara},
  Y., {Ketzeback}, W., {Kollatschny}, W., {Konishi}, K., {Marshall}, J.~L.,
  {McMillan}, R.~J., {Miknaitis}, G., {Morokuma}, T., {M{\"o}rtsell}, E.,
  {Pan}, K., {Prieto}, J.~L., {Richmond}, M.~W., {Riess}, A.~G., {Romani}, R.,
  {Schneider}, D.~P., {Sollerman}, J., {Takanashi}, N., {Tokita}, K., {van der
  Heyden}, K., {Wheeler}, J.~C., {Yasuda}, N., \& {York}, D. 2009, \apjs, 185,
  32

\bibitem[{{Komatsu} {et~al.}(2011){Komatsu}, {Smith}, {Dunkley}, {Bennett},
  {Gold}, {Hinshaw}, {Jarosik}, {Larson}, {Nolta}, {Page}, {Spergel},
  {Halpern}, {Hill}, {Kogut}, {Limon}, {Meyer}, {Odegard}, {Tucker}, {Weiland},
  {Wollack}, \& {Wright}}]{komatsu11}
{Komatsu}, E., {Smith}, K.~M., {Dunkley}, J., {Bennett}, C.~L., {Gold}, B.,
  {Hinshaw}, G., {Jarosik}, N., {Larson}, D., {Nolta}, M.~R., {Page}, L.,
  {Spergel}, D.~N., {Halpern}, M., {Hill}, R.~S., {Kogut}, A., {Limon}, M.,
  {Meyer}, S.~S., {Odegard}, N., {Tucker}, G.~S., {Weiland}, J.~L., {Wollack},
  E., \& {Wright}, E.~L. 2011, \apjs, "in press"

\bibitem[{{Kunz} {et~al.}(2007){Kunz}, {Bassett}, \& {Hlozek}}]{kunz07}
{Kunz}, M., {Bassett}, B.~A., \& {Hlozek}, R.~A. 2007, \prd, 75, 103508

\bibitem[{{Lampeitl} {et~al.}(2010){Lampeitl}, {Smith}, {Nichol}, {Bassett},
  {Cinabro}, {Dilday}, {Foley}, {Frieman}, {Garnavich}, {Goobar}, {Im}, {Jha},
  {Marriner}, {Miquel}, {Nordin}, {{\"O}stman}, {Riess}, {Sako}, {Schneider},
  {Sollerman}, \& {Stritzinger}}]{lampeitl10}
{Lampeitl}, H., {Smith}, M., {Nichol}, R.~C., {Bassett}, B., {Cinabro}, D.,
  {Dilday}, B., {Foley}, R.~J., {Frieman}, J.~A., {Garnavich}, P.~M., {Goobar},
  A., {Im}, M., {Jha}, S.~W., {Marriner}, J., {Miquel}, R., {Nordin}, J.,
  {{\"O}stman}, L., {Riess}, A.~G., {Sako}, M., {Schneider}, D.~P.,
  {Sollerman}, J., \& {Stritzinger}, M. 2010, \apj, 722, 566

\bibitem[{{Macri} {et~al.}(2011{\natexlab{a}}){Macri}, {Riess}, {Casertano},
  {Lampeitl}, {Ferguson}, {Filippenko}, {Jha}, {Li}, \& {Chornock}}]{macri11c}
{Macri}, L.~M., {Riess}, A.~G., {Casertano}, S., {Lampeitl}, H., {Ferguson},
  H.~C., {Filippenko}, A.~V., {Jha}, S.~W., {Li}, W., \& {Chornock}, R.
  2011{\natexlab{a}}, \apjs, in prep.

\bibitem[{{Macri} {et~al.}(2011{\natexlab{b}}){Macri}, {Riess}, {Casertano},
  {Lampeitl}, {Ferguson}, {Filippenko}, {Jha}, {Li}, {Chornock}, \&
  {Welch}}]{macri11a}
{Macri}, L.~M., {Riess}, A.~G., {Casertano}, S., {Lampeitl}, H., {Ferguson},
  H.~C., {Filippenko}, A.~V., {Jha}, S.~W., {Li}, W., {Chornock}, R., \&
  {Welch}, D.~L. 2011{\natexlab{b}}, \apj, in prep.

\bibitem[{{Macri} {et~al.}(2011{\natexlab{c}}){Macri}, {Riess}, {Casertano},
  {Lampeitl}, {Ferguson}, {Filippenko}, {Jha}, {Li}, {Chornock}, {Welch}, \&
  {Withmore}}]{macri11b}
{Macri}, L.~M., {Riess}, A.~G., {Casertano}, S., {Lampeitl}, H., {Ferguson},
  H.~C., {Filippenko}, A.~V., {Jha}, S.~W., {Li}, W., {Chornock}, R., {Welch},
  D.~L., \& {Withmore}, B.~C. 2011{\natexlab{c}}, \apj, in prep.

\bibitem[{{Macri} {et~al.}(2006){Macri}, {Stanek}, {Bersier}, {Greenhill}, \&
  {Reid}}]{macri06}
{Macri}, L.~M., {Stanek}, K.~Z., {Bersier}, D., {Greenhill}, L.~J., \& {Reid},
  M.~J. 2006, \apj, 652, 1133

\bibitem[{{Madore}(1982)}]{madore82}
{Madore}, B.~F. 1982, \apj, 253, 575

\bibitem[{{Madore} \& {Freedman}(1991)}]{madore91}
{Madore}, B.~F. \& {Freedman}, W.~L. 1991, \pasp, 103, 933

\bibitem[{{Mangano} \& {Serpico}(2005)}]{mangano05}
{Mangano}, G. \& {Serpico}, P.~D. 2005, Nuclear Physics B Proceedings
  Supplements, 145, 351

\bibitem[{{Marconi} {et~al.}(2005){Marconi}, {Musella}, \&
  {Fiorentino}}]{marconi05}
{Marconi}, M., {Musella}, I., \& {Fiorentino}, G. 2005, \apj, 632, 590

\bibitem[{{McGaugh}(1991)}]{mcgaugh91}
{McGaugh}, S.~S. 1991, \apj, 380, 140

\bibitem[{{Nadathur} \& {Sarkar}(2010)}]{nadathur10}
{Nadathur}, S. \& {Sarkar}, S. 2010, ArXiv e-prints

\bibitem[{{Neill} {et~al.}(2007){Neill}, {Hudson}, \& {Conley}}]{neill07}
{Neill}, J.~D., {Hudson}, M.~J., \& {Conley}, A. 2007, \apjl, 661, L123

\bibitem[{{Neill} {et~al.}(2009){Neill}, {Sullivan}, {Howell}, {Conley},
  {Seibert}, {Martin}, {Barlow}, {Foster}, {Friedman}, {Morrissey}, {Neff},
  {Schiminovich}, {Wyder}, {Bianchi}, {Donas}, {Heckman}, {Lee}, {Madore},
  {Milliard}, {Rich}, \& {Szalay}}]{neill09}
{Neill}, J.~D., {Sullivan}, M., {Howell}, D.~A., {Conley}, A., {Seibert}, M.,
  {Martin}, D.~C., {Barlow}, T.~A., {Foster}, K., {Friedman}, P.~G.,
  {Morrissey}, P., {Neff}, S.~G., {Schiminovich}, D., {Wyder}, T.~K.,
  {Bianchi}, L., {Donas}, J., {Heckman}, T.~M., {Lee}, Y., {Madore}, B.~F.,
  {Milliard}, B., {Rich}, R.~M., \& {Szalay}, A.~S. 2009, \apj, 707, 1449

\bibitem[{{Perlmutter} {et~al.}(1999){Perlmutter}, {Aldering}, {Goldhaber},
  {Knop}, {Nugent}, {Castro}, {Deustua}, {Fabbro}, {Goobar}, {Groom}, {Hook},
  {Kim}, {Kim}, {Lee}, {Nunes}, {Pain}, {Pennypacker}, {Quimby}, {Lidman},
  {Ellis}, {Irwin}, {McMahon}, {Ruiz-Lapuente}, {Walton}, {Schaefer}, {Boyle},
  {Filippenko}, {Matheson}, {Fruchter}, {Panagia}, {Newberg}, {Couch}, \& {The
  Supernova Cosmology Project}}]{perlmutter99}
{Perlmutter}, S., {Aldering}, G., {Goldhaber}, G., {Knop}, R.~A., {Nugent}, P.,
  {Castro}, P.~G., {Deustua}, S., {Fabbro}, S., {Goobar}, A., {Groom}, D.~E.,
  {Hook}, I.~M., {Kim}, A.~G., {Kim}, M.~Y., {Lee}, J.~C., {Nunes}, N.~J.,
  {Pain}, R., {Pennypacker}, C.~R., {Quimby}, R., {Lidman}, C., {Ellis}, R.~S.,
  {Irwin}, M., {McMahon}, R.~G., {Ruiz-Lapuente}, P., {Walton}, N., {Schaefer},
  B., {Boyle}, B.~J., {Filippenko}, A.~V., {Matheson}, T., {Fruchter}, A.~S.,
  {Panagia}, N., {Newberg}, H.~J.~M., {Couch}, W.~J., \& {The Supernova
  Cosmology Project}. 1999, \apj, 517, 565

\bibitem[{{Persson} {et~al.}(2004){Persson}, {Madore}, {Krzemi{\'n}ski},
  {Freedman}, {Roth}, \& {Murphy}}]{persson04}
{Persson}, S.~E., {Madore}, B.~F., {Krzemi{\'n}ski}, W., {Freedman}, W.~L.,
  {Roth}, M., \& {Murphy}, D.~C. 2004, \aj, 128, 2239

\bibitem[{{Persson} {et~al.}(1998){Persson}, {Murphy}, {Krzeminski}, {Roth}, \&
  {Rieke}}]{persson98}
{Persson}, S.~E., {Murphy}, D.~C., {Krzeminski}, W., {Roth}, M., \& {Rieke},
  M.~J. 1998, \aj, 116, 2475

\bibitem[{{Pietrzy{\'n}ski} {et~al.}(2009){Pietrzy{\'n}ski}, {Thompson},
  {Graczyk}, {Gieren}, {Udalski}, {Szewczyk}, {Minniti}, {Ko{\l}aczkowski},
  {Bresolin}, \& {Kudritzki}}]{pietrzynski09}
{Pietrzy{\'n}ski}, G., {Thompson}, I.~B., {Graczyk}, D., {Gieren}, W.,
  {Udalski}, A., {Szewczyk}, O., {Minniti}, D., {Ko{\l}aczkowski}, Z.,
  {Bresolin}, F., \& {Kudritzki}, R. 2009, \apj, 697, 862

\bibitem[{{Pike} \& {Hudson}(2005)}]{pike05}
{Pike}, R.~W. \& {Hudson}, M.~J. 2005, \apj, 635, 11

\bibitem[{{Pilyugin} \& {Thuan}(2005)}]{pilyugin05}
{Pilyugin}, L.~S. \& {Thuan}, T.~X. 2005, \apj, 631, 231

\bibitem[{{Reid} {et~al.}(2010){Reid}, {Verde}, {Jimenez}, \& {Mena}}]{reid10}
{Reid}, B.~A., {Verde}, L., {Jimenez}, R., \& {Mena}, O. 2010, \jcap, 1, 3

\bibitem[{{Ribas} {et~al.}(2002){Ribas}, {Fitzpatrick}, {Maloney}, {Guinan}, \&
  {Udalski}}]{ribas02}
{Ribas}, I., {Fitzpatrick}, E.~L., {Maloney}, F.~P., {Guinan}, E.~F., \&
  {Udalski}, A. 2002, \apj, 574, 771

\bibitem[{{Riess} {et~al.}(1997){Riess}, {Davis}, {Baker}, \&
  {Kirshner}}]{riess97}
{Riess}, A.~G., {Davis}, M., {Baker}, J., \& {Kirshner}, R.~P. 1997, \apjl,
  488, L1+

\bibitem[{{Riess} {et~al.}(1998){Riess}, {Filippenko}, {Challis},
  {Clocchiatti}, {Diercks}, {Garnavich}, {Gilliland}, {Hogan}, {Jha},
  {Kirshner}, {Leibundgut}, {Phillips}, {Reiss}, {Schmidt}, {Schommer},
  {Smith}, {Spyromilio}, {Stubbs}, {Suntzeff}, \& {Tonry}}]{riess98}
{Riess}, A.~G., {Filippenko}, A.~V., {Challis}, P., {Clocchiatti}, A.,
  {Diercks}, A., {Garnavich}, P.~M., {Gilliland}, R.~L., {Hogan}, C.~J., {Jha},
  S., {Kirshner}, R.~P., {Leibundgut}, B., {Phillips}, M.~M., {Reiss}, D.,
  {Schmidt}, B.~P., {Schommer}, R.~A., {Smith}, R.~C., {Spyromilio}, J.,
  {Stubbs}, C., {Suntzeff}, N.~B., \& {Tonry}, J. 1998, \aj, 116, 1009

\bibitem[{{Riess} {et~al.}(2005){Riess}, {Li}, {Stetson}, {Filippenko}, {Jha},
  {Kirshner}, {Challis}, {Garnavich}, \& {Chornock}}]{riess05}
{Riess}, A.~G., {Li}, W., {Stetson}, P.~B., {Filippenko}, A.~V., {Jha}, S.,
  {Kirshner}, R.~P., {Challis}, P.~M., {Garnavich}, P.~M., \& {Chornock}, R.
  2005, \apj, 627, 579

\bibitem[{{Riess} {et~al.}(2009{\natexlab{a}}){Riess}, {Macri}, {Casertano},
  {Sosey}, {Lampeitl}, {Ferguson}, {Filippenko}, {Jha}, {Li}, {Chornock}, \&
  {Sarkar}}]{riess09a}
{Riess}, A.~G., {Macri}, L., {Casertano}, S., {Sosey}, M., {Lampeitl}, H.,
  {Ferguson}, H.~C., {Filippenko}, A.~V., {Jha}, S.~W., {Li}, W., {Chornock},
  R., \& {Sarkar}, D. 2009{\natexlab{a}}, \apj, 699, 539

\bibitem[{{Riess} {et~al.}(2009{\natexlab{b}}){Riess}, {Macri}, {Li},
  {Lampeitl}, {Casertano}, {Ferguson}, {Filippenko}, {Jha}, {Chornock},
  {Greenhill}, {Mutchler}, {Ganeshalingham}, \& {Hicken}}]{riess09b}
{Riess}, A.~G., {Macri}, L., {Li}, W., {Lampeitl}, H., {Casertano}, S.,
  {Ferguson}, H.~C., {Filippenko}, A.~V., {Jha}, S.~W., {Chornock}, R.,
  {Greenhill}, L., {Mutchler}, M., {Ganeshalingham}, M., \& {Hicken}, M.
  2009{\natexlab{b}}, \apjs, 183, 109

\bibitem[{{Riess} {et~al.}(2007){Riess}, {Strolger}, {Casertano}, {Ferguson},
  {Mobasher}, {Gold}, {Challis}, {Filippenko}, {Jha}, {Li}, {Tonry}, {Foley},
  {Kirshner}, {Dickinson}, {MacDonald}, {Eisenstein}, {Livio}, {Younger}, {Xu},
  {Dahl{\'e}n}, \& {Stern}}]{riess07}
{Riess}, A.~G., {Strolger}, L., {Casertano}, S., {Ferguson}, H.~C., {Mobasher},
  B., {Gold}, B., {Challis}, P.~J., {Filippenko}, A.~V., {Jha}, S., {Li}, W.,
  {Tonry}, J., {Foley}, R., {Kirshner}, R.~P., {Dickinson}, M., {MacDonald},
  E., {Eisenstein}, D., {Livio}, M., {Younger}, J., {Xu}, C., {Dahl{\'e}n}, T.,
  \& {Stern}, D. 2007, \apj, 659, 98

\bibitem[{{Romaniello} {et~al.}(2008){Romaniello}, {Primas}, {Mottini},
  {Pedicelli}, {Lemasle}, {Bono}, {Fran{\c c}ois}, {Groenewegen}, \&
  {Laney}}]{romaniello08}
{Romaniello}, M., {Primas}, F., {Mottini}, M., {Pedicelli}, S., {Lemasle}, B.,
  {Bono}, G., {Fran{\c c}ois}, P., {Groenewegen}, M.~A.~T., \& {Laney}, C.~D.
  2008, \aap, 488, 731

\bibitem[{{Saha} {et~al.}(1996){Saha}, {Sandage}, {Labhardt}, {Tammann},
  {Macchetto}, \& {Panagia}}]{saha96}
{Saha}, A., {Sandage}, A., {Labhardt}, L., {Tammann}, G.~A., {Macchetto},
  F.~D., \& {Panagia}, N. 1996, \apj, 466, 55

\bibitem[{{Saha} {et~al.}(1997){Saha}, {Sandage}, {Labhardt}, {Tammann},
  {Macchetto}, \& {Panagia}}]{saha97}
---. 1997, \apj, 486, 1

\bibitem[{{Saha} {et~al.}(2001){Saha}, {Sandage}, {Tammann}, {Dolphin},
  {Christensen}, {Panagia}, \& {Macchetto}}]{saha01}
{Saha}, A., {Sandage}, A., {Tammann}, G.~A., {Dolphin}, A.~E., {Christensen},
  J., {Panagia}, N., \& {Macchetto}, F.~D. 2001, \apj, 562, 314

\bibitem[{{Sakai} {et~al.}(2004){Sakai}, {Ferrarese}, {Kennicutt}, \&
  {Saha}}]{sakai04}
{Sakai}, S., {Ferrarese}, L., {Kennicutt}, Jr., R.~C., \& {Saha}, A. 2004,
  \apj, 608, 42

\bibitem[{{Sandage} {et~al.}(2006){Sandage}, {Tammann}, {Saha}, {Reindl},
  {Macchetto}, \& {Panagia}}]{sandage06}
{Sandage}, A., {Tammann}, G.~A., {Saha}, A., {Reindl}, B., {Macchetto}, F.~D.,
  \& {Panagia}, N. 2006, \apj, 653, 843

\bibitem[{{Saviane} {et~al.}(2008){Saviane}, {Momany}, {da Costa}, {Rich}, \&
  {Hibbard}}]{saviane08}
{Saviane}, I., {Momany}, Y., {da Costa}, G.~S., {Rich}, R.~M., \& {Hibbard},
  J.~E. 2008, \apj, 678, 179

\bibitem[{{Schaefer}(2008)}]{schaefer08}
{Schaefer}, B.~E. 2008, \aj, 135, 112

\bibitem[{{Schweizer} {et~al.}(2008){Schweizer}, {Burns}, {Madore}, {Mager},
  {Phillips}, {Freedman}, {Boldt}, {Contreras}, {Folatelli}, {Gonz{\'a}lez},
  {Hamuy}, {Krzeminski}, {Morrell}, {Persson}, {Roth}, \&
  {Stritzinger}}]{schweizer08}
{Schweizer}, F., {Burns}, C.~R., {Madore}, B.~F., {Mager}, V.~A., {Phillips},
  M.~M., {Freedman}, W.~L., {Boldt}, L., {Contreras}, C., {Folatelli}, G.,
  {Gonz{\'a}lez}, S., {Hamuy}, M., {Krzeminski}, W., {Morrell}, N.~I.,
  {Persson}, S.~E., {Roth}, M.~R., \& {Stritzinger}, M.~D. 2008, \aj, 136, 1482

\bibitem[{{Skrutskie} {et~al.}(2006){Skrutskie}, {Cutri}, {Stiening},
  {Weinberg}, {Schneider}, {Carpenter}, {Beichman}, {Capps}, {Chester},
  {Elias}, {Huchra}, {Liebert}, {Lonsdale}, {Monet}, {Price}, {Seitzer},
  {Jarrett}, {Kirkpatrick}, {Gizis}, {Howard}, {Evans}, {Fowler}, {Fullmer},
  {Hurt}, {Light}, {Kopan}, {Marsh}, {McCallon}, {Tam}, {Van Dyk}, \&
  {Wheelock}}]{skrutskie06}
{Skrutskie}, M.~F., {Cutri}, R.~M., {Stiening}, R., {Weinberg}, M.~D.,
  {Schneider}, S., {Carpenter}, J.~M., {Beichman}, C., {Capps}, R., {Chester},
  T., {Elias}, J., {Huchra}, J., {Liebert}, J., {Lonsdale}, C., {Monet}, D.~G.,
  {Price}, S., {Seitzer}, P., {Jarrett}, T., {Kirkpatrick}, J.~D., {Gizis},
  J.~E., {Howard}, E., {Evans}, T., {Fowler}, J., {Fullmer}, L., {Hurt}, R.,
  {Light}, R., {Kopan}, E.~L., {Marsh}, K.~A., {McCallon}, H.~L., {Tam}, R.,
  {Van Dyk}, S., \& {Wheelock}, S. 2006, \aj, 131, 1163

\bibitem[{{Soszy{\'n}ski} {et~al.}(2005){Soszy{\'n}ski}, {Gieren}, \&
  {Pietrzy{\'n}ski}}]{soszynski05}
{Soszy{\'n}ski}, I., {Gieren}, W., \& {Pietrzy{\'n}ski}, G. 2005, \pasp, 117,
  823

\bibitem[{{Stetson} \& {Gibson}(2001)}]{stetson01}
{Stetson}, P.~B. \& {Gibson}, B.~K. 2001, \mnras, 328, L1

\bibitem[{{Sullivan} {et~al.}(2010){Sullivan}, {Conley}, {Howell}, {Neill},
  {Astier}, {Balland}, {Basa}, {Carlberg}, {Fouchez}, {Guy}, {Hardin}, {Hook},
  {Pain}, {Palanque-Delabrouille}, {Perrett}, {Pritchet}, {Regnault}, {Rich},
  {Ruhlmann-Kleider}, {Baumont}, {Hsiao}, {Kronborg}, {Lidman}, {Perlmutter},
  \& {Walker}}]{sullivan10}
{Sullivan}, M., {Conley}, A., {Howell}, D.~A., {Neill}, J.~D., {Astier}, P.,
  {Balland}, C., {Basa}, S., {Carlberg}, R.~G., {Fouchez}, D., {Guy}, J.,
  {Hardin}, D., {Hook}, I.~M., {Pain}, R., {Palanque-Delabrouille}, N.,
  {Perrett}, K.~M., {Pritchet}, C.~J., {Regnault}, N., {Rich}, J.,
  {Ruhlmann-Kleider}, V., {Baumont}, S., {Hsiao}, E., {Kronborg}, T., {Lidman},
  C., {Perlmutter}, S., \& {Walker}, E.~S. 2010, \mnras, 406, 782

\bibitem[{{Tonry} {et~al.}(2000){Tonry}, {Blakeslee}, {Ajhar}, \&
  {Dressler}}]{tonry00}
{Tonry}, J.~L., {Blakeslee}, J.~P., {Ajhar}, E.~A., \& {Dressler}, A. 2000,
  \apj, 530, 625

\bibitem[{{Udalski} {et~al.}(1999){Udalski}, {Soszynski}, {Szymanski},
  {Kubiak}, {Pietrzynski}, {Wozniak}, \& {Zebrun}}]{udalski99}
{Udalski}, A., {Soszynski}, I., {Szymanski}, M., {Kubiak}, M., {Pietrzynski},
  G., {Wozniak}, P., \& {Zebrun}, K. 1999, \actaa, 49, 223

\bibitem[{{van Leeuwen} {et~al.}(2007){van Leeuwen}, {Feast}, {Whitelock}, \&
  {Laney}}]{vanleeuwen07}
{van Leeuwen}, F., {Feast}, M.~W., {Whitelock}, P.~A., \& {Laney}, C.~D. 2007,
  \mnras, 379, 723

\bibitem[{{Whitmore} {et~al.}(1999){Whitmore}, {Zhang}, {Leitherer}, {Fall},
  {Schweizer}, \& {Miller}}]{whitmore99}
{Whitmore}, B.~C., {Zhang}, Q., {Leitherer}, C., {Fall}, S.~M., {Schweizer},
  F., \& {Miller}, B.~W. 1999, \aj, 118, 1551

\bibitem[{{Wiltshire}(2007)}]{wiltshire07}
{Wiltshire}, D.~L. 2007, Physical Review Letters, 99, 251101

\bibitem[{{Zaritsky} {et~al.}(1994){Zaritsky}, {Kennicutt}, \&
  {Huchra}}]{zaritsky94}
{Zaritsky}, D., {Kennicutt}, Jr., R.~C., \& {Huchra}, J.~P. 1994, \apj, 420, 87

\end{thebibliography}

\tabletypesize{\scriptsize}
\tablewidth{0pt}


\begin{figure}[ht]
\vspace*{140mm}
\figurenum{1}
\includegraphics{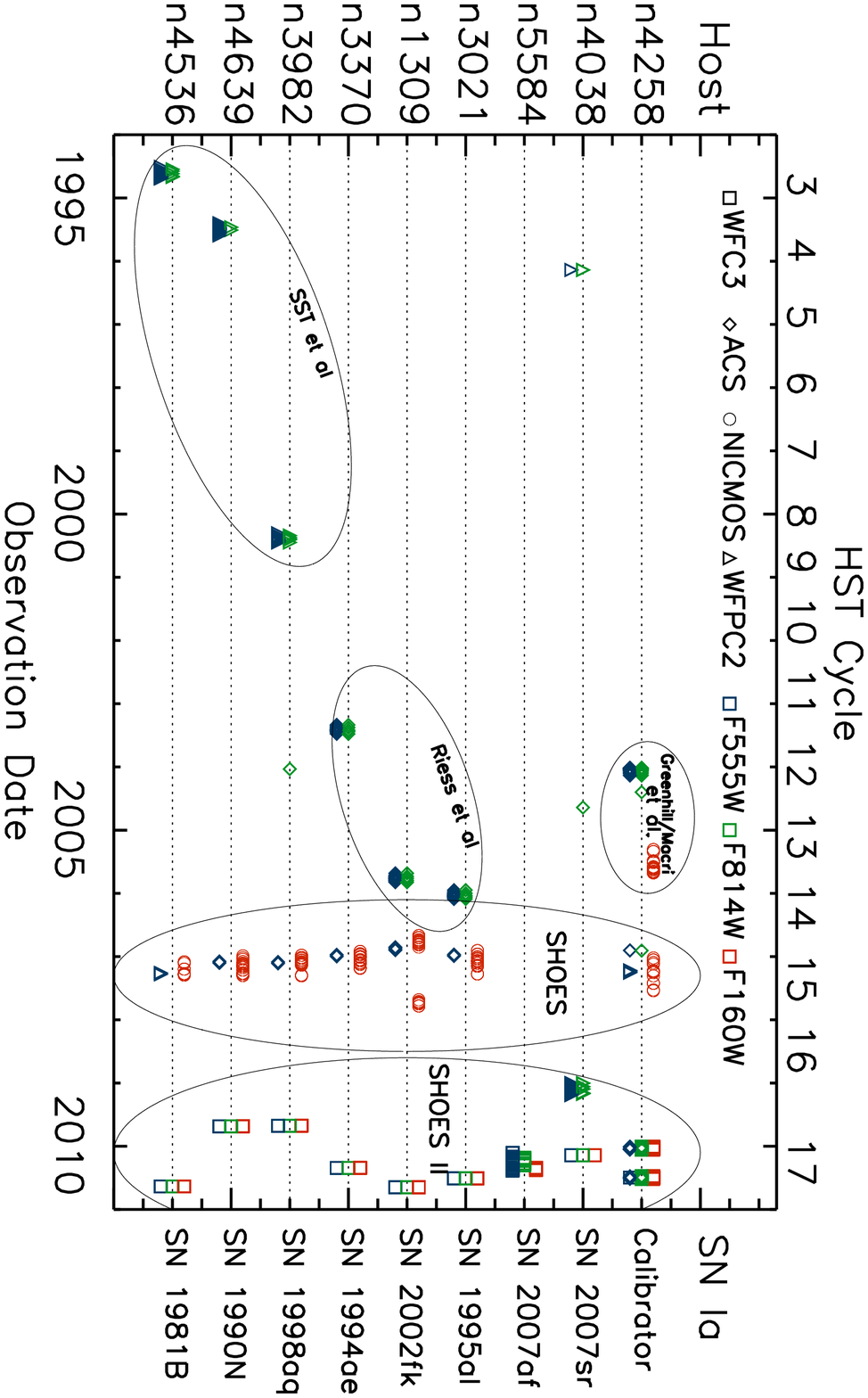}
\caption { }
\end{figure}

\begin{figure}[ht]
\vspace*{140mm}
\figurenum{2}
\includegraphics{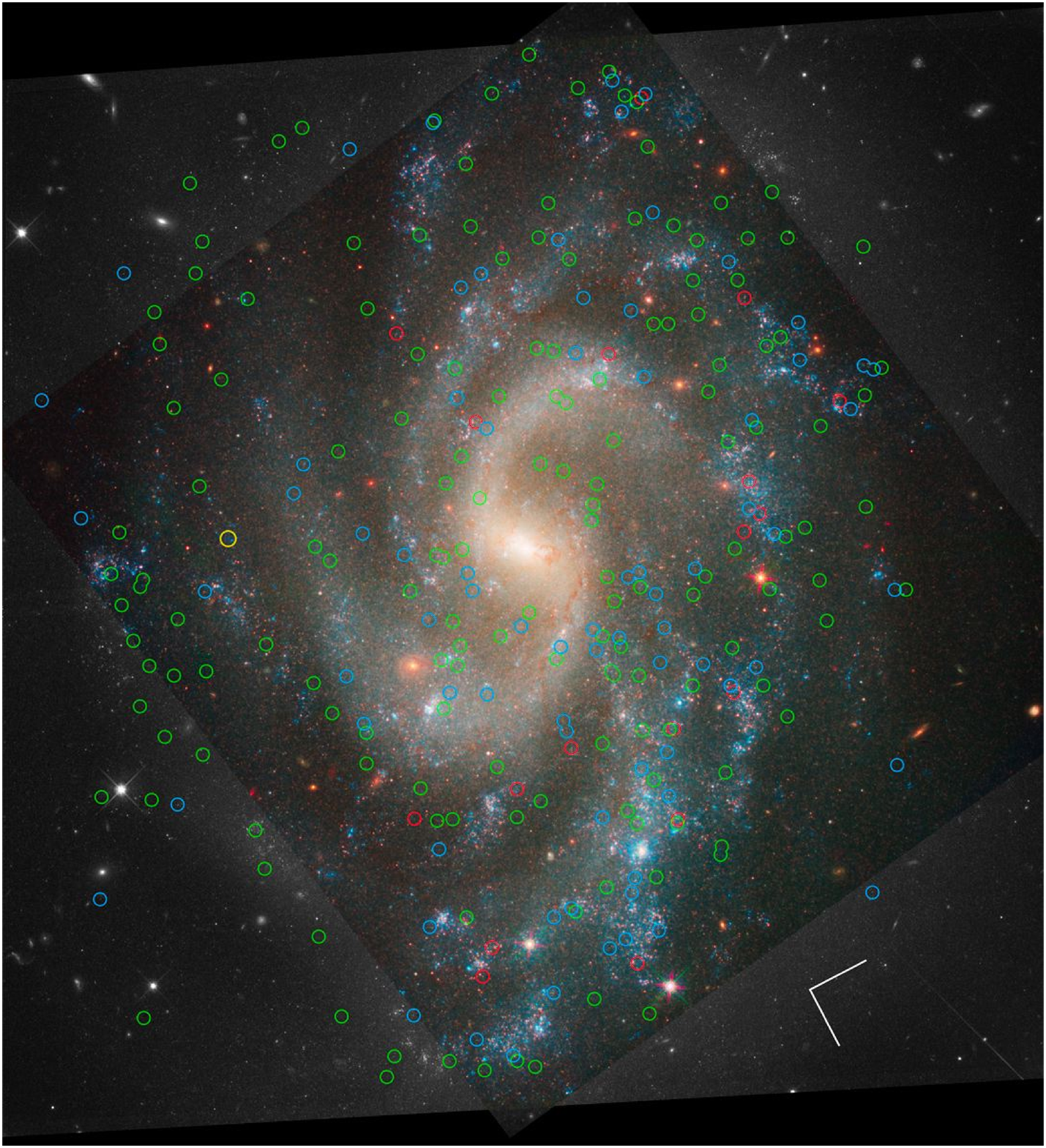}
\caption { }
\end{figure}

\begin{figure}[ht]
\vspace*{140mm}
\figurenum{3}
\includegraphics{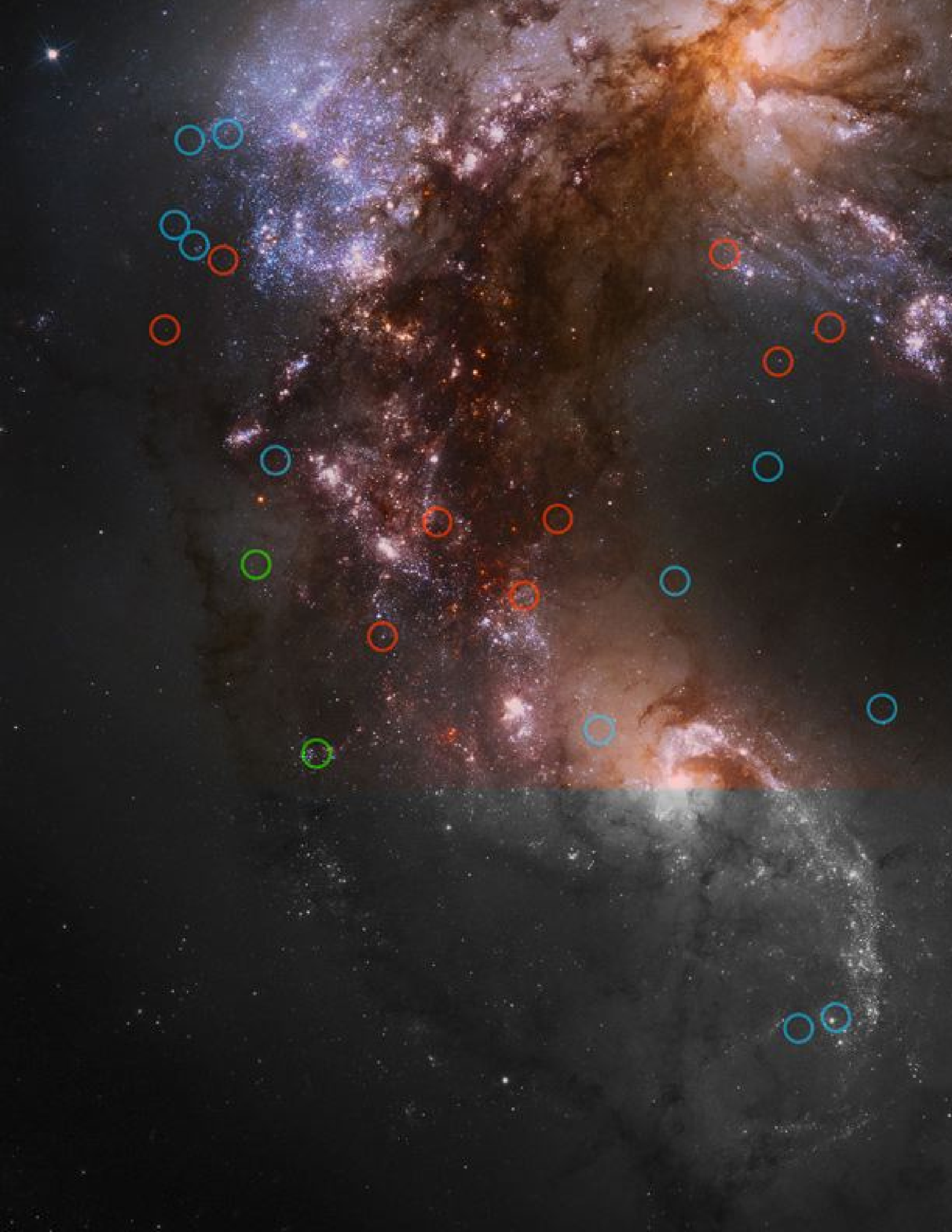}
\caption { }
\end{figure}

\begin{figure}[ht]
\vspace*{140mm}
\figurenum{4a}
\includegraphics{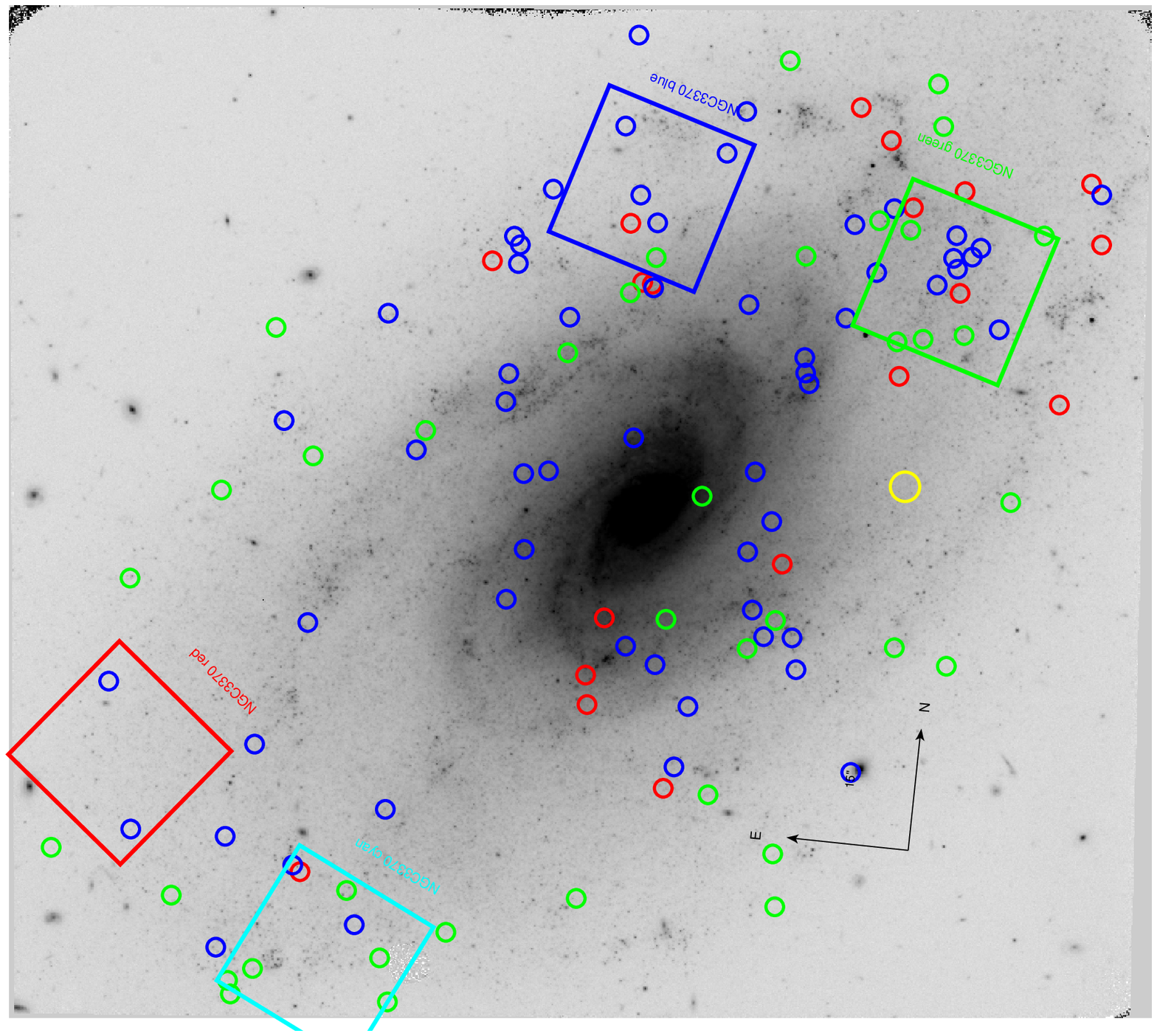}
\caption { }
\end{figure}

\begin{figure}[ht]
\vspace*{140mm}
\figurenum{4b}
\includegraphics{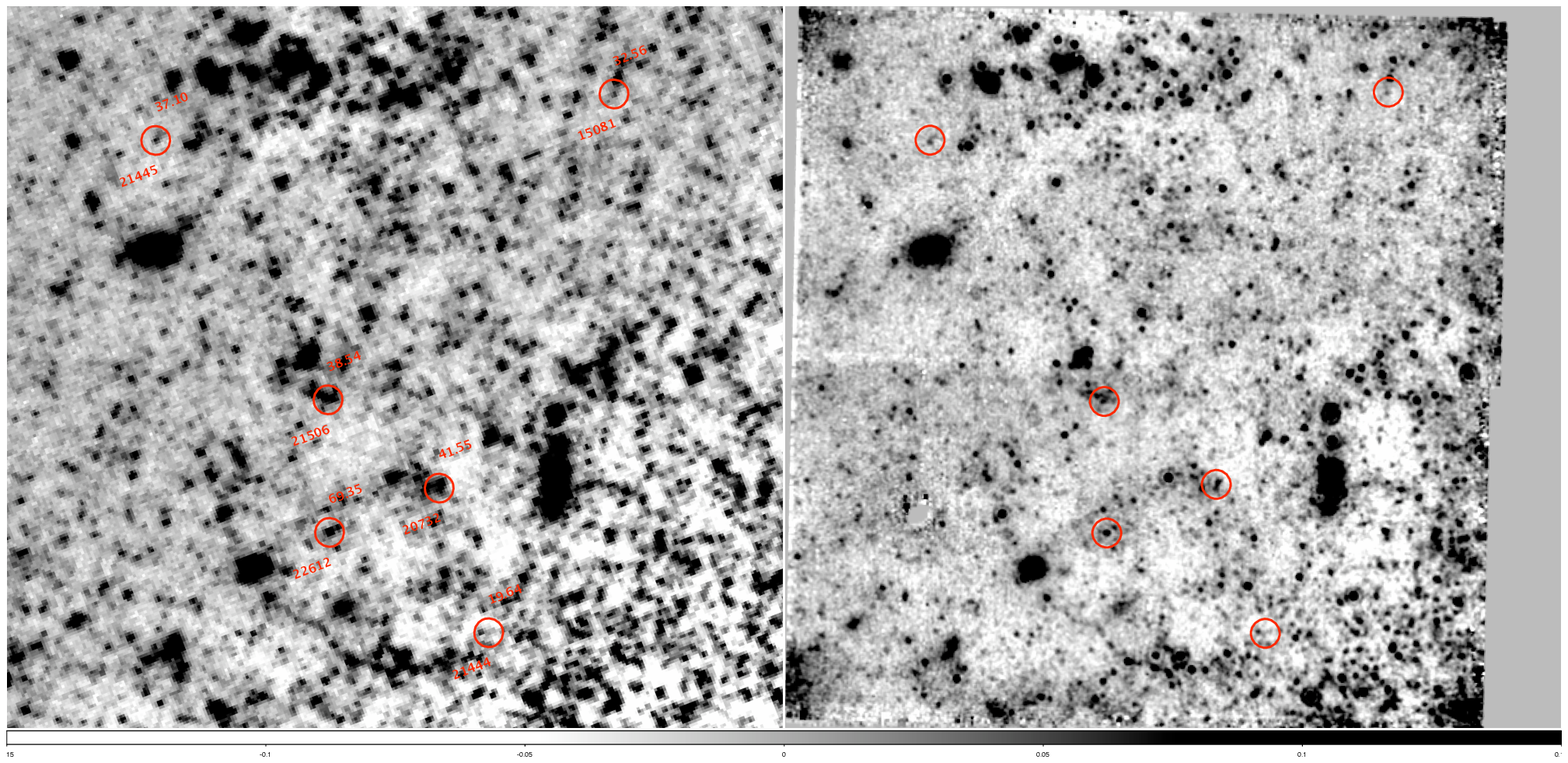}
\caption { }
\end{figure}

\begin{figure}[ht]
\vspace*{140mm}
\figurenum{5}
\includegraphics{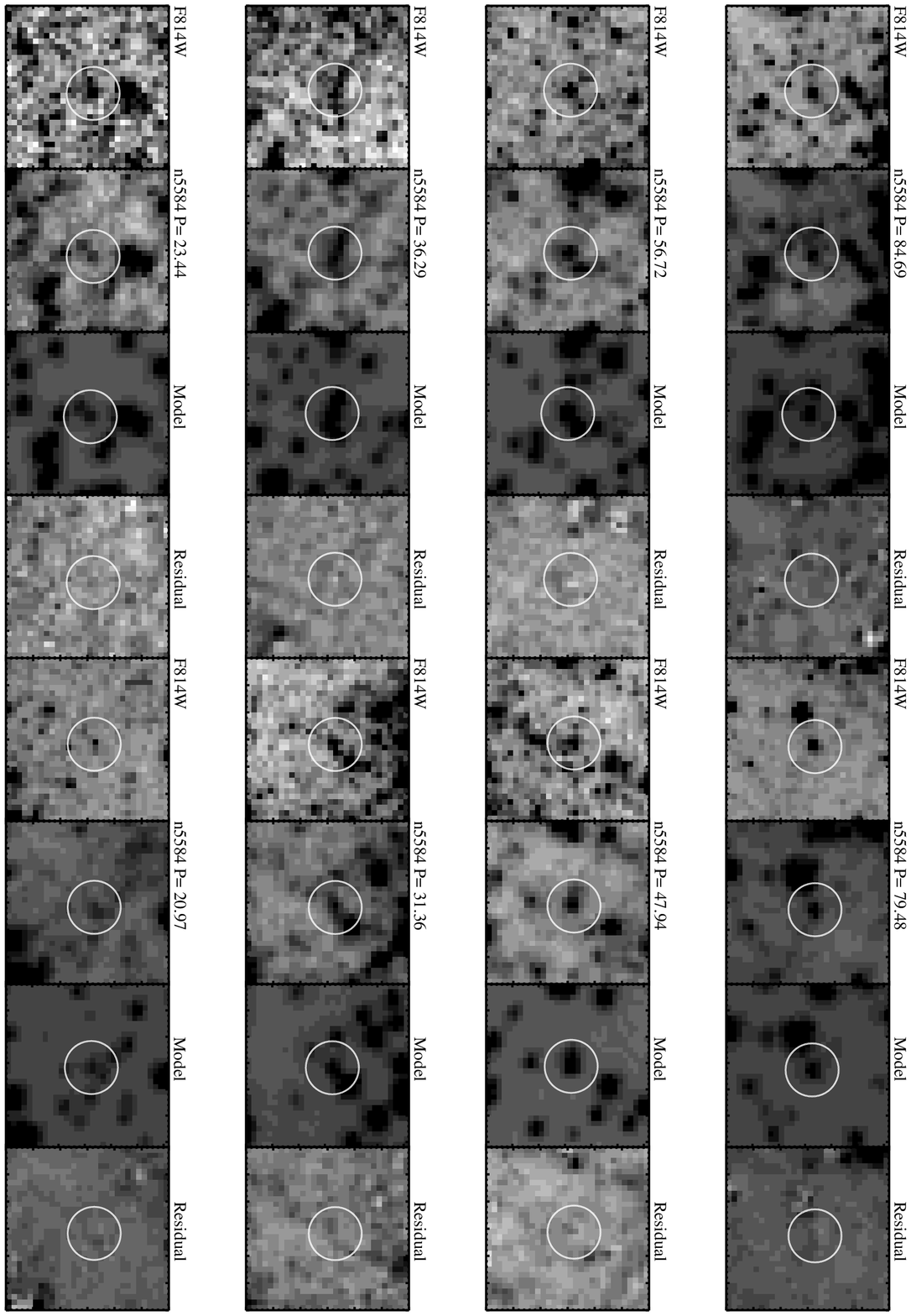}
\caption { }
\end{figure}

% generated with crowd_wfc4_all_5584_examp in irphot

\begin{figure}[ht]
\vspace*{140mm}
\figurenum{6}
\includegraphics{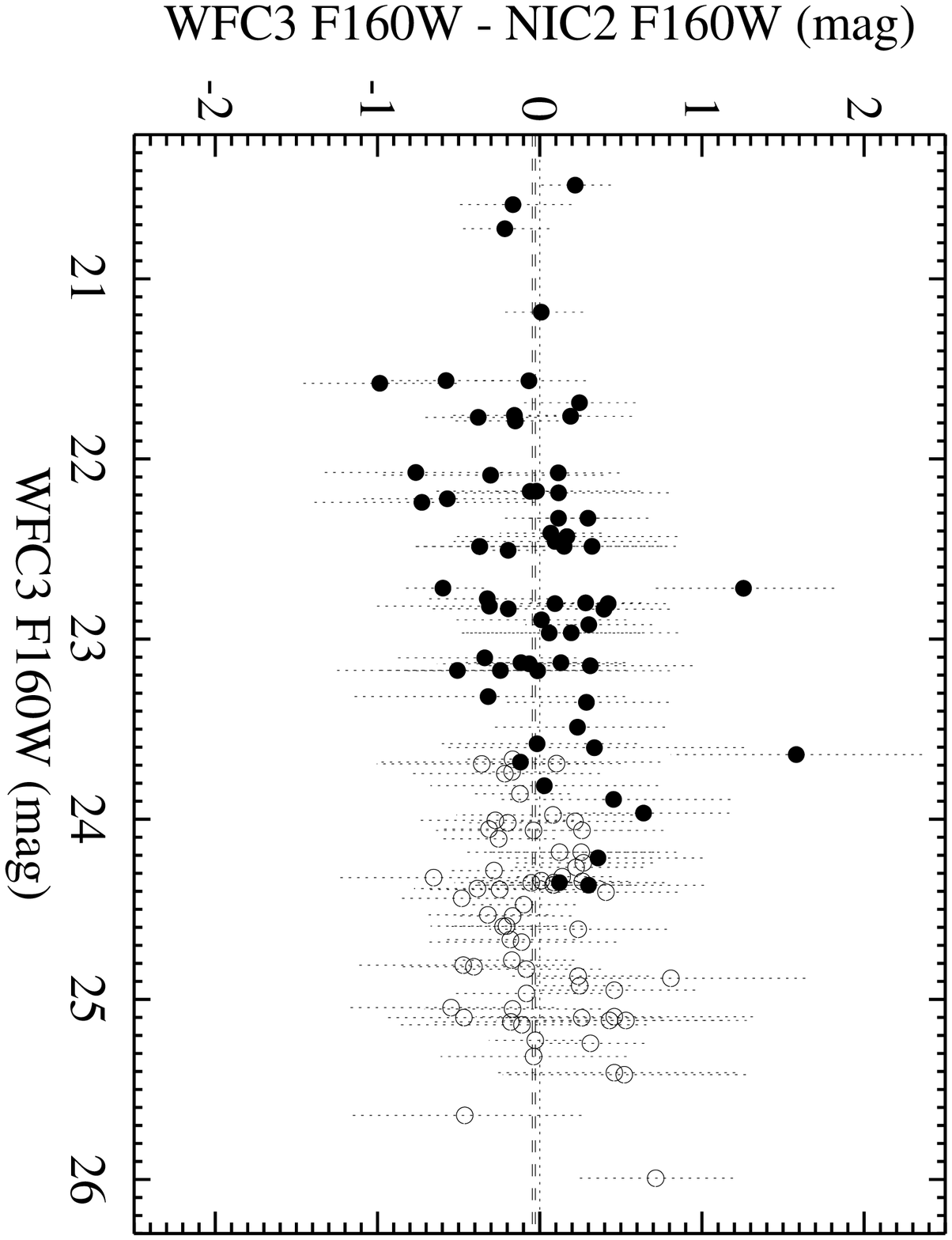}
\caption { }
\end{figure}

\begin{figure}[ht]
\vspace*{140mm}
\figurenum{7}
\includegraphics{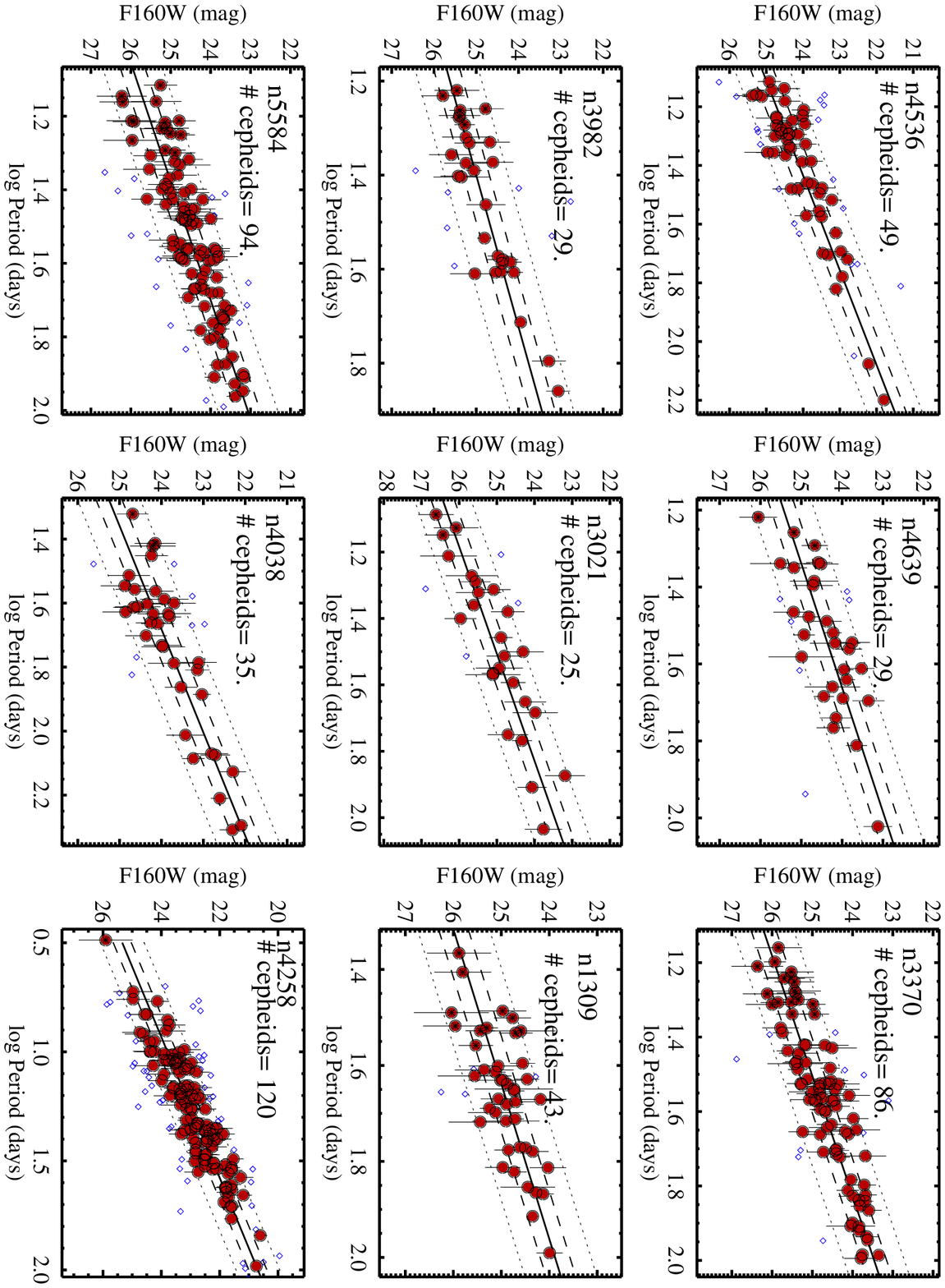}
\caption { }
\end{figure}

\begin{figure}[ht]
\vspace*{140mm}
\figurenum{8}
\includegraphics{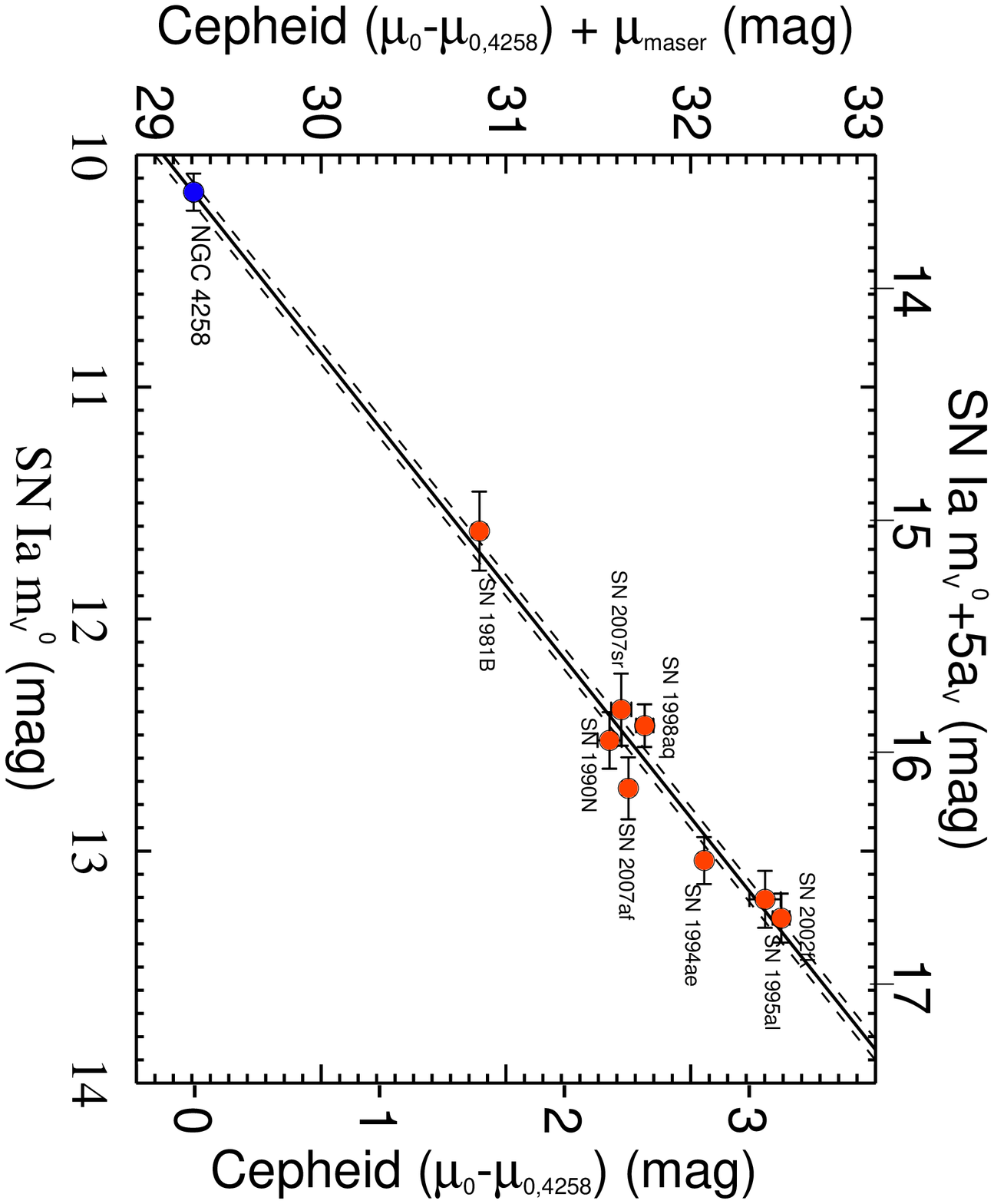}
\caption { }
\end{figure}

\begin{figure}[ht]
\vspace*{140mm}
\figurenum{9}
\includegraphics{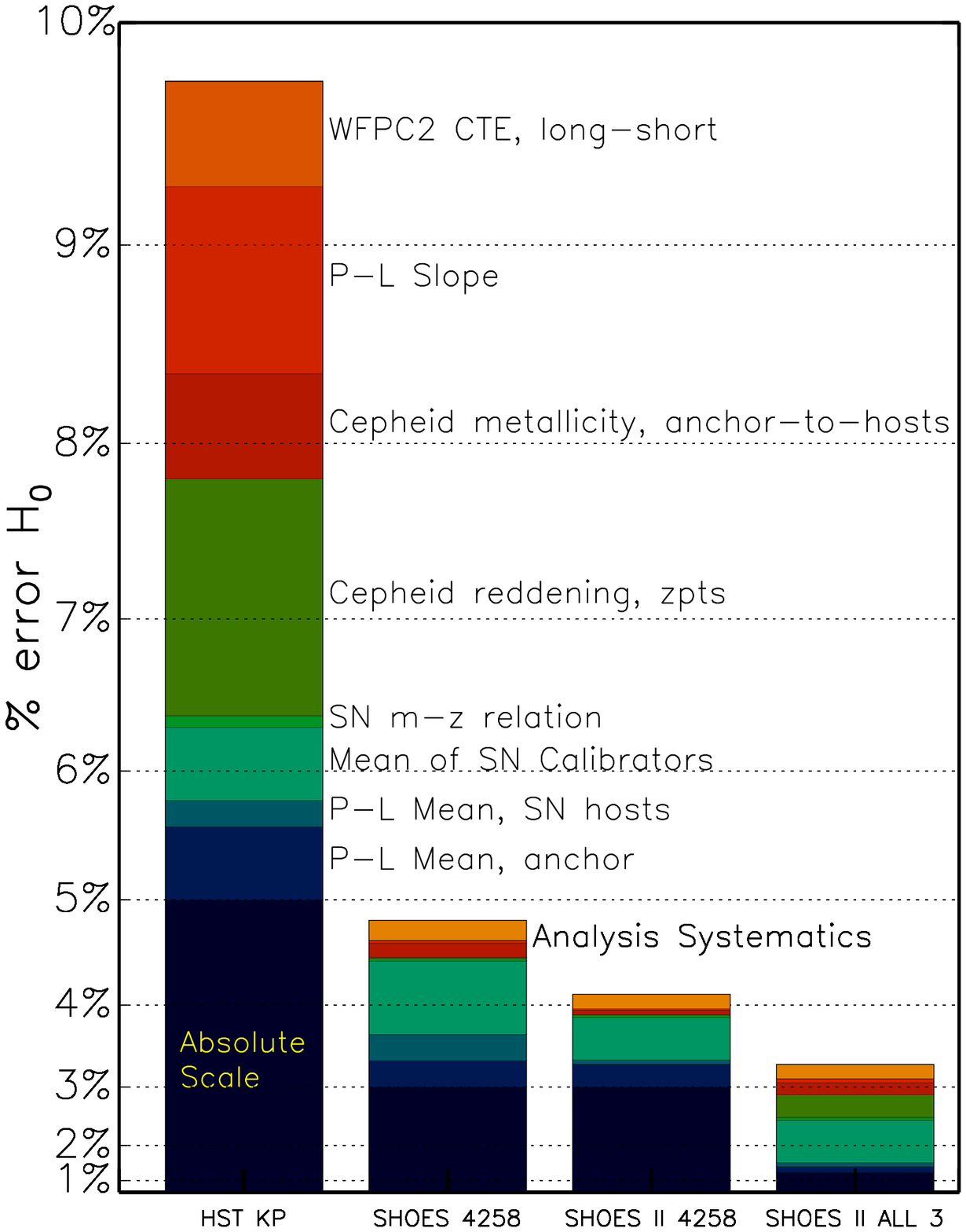}
\caption { }
\end{figure}

\begin{figure}[ht]
\vspace*{140mm}
\figurenum{10}
\includegraphics{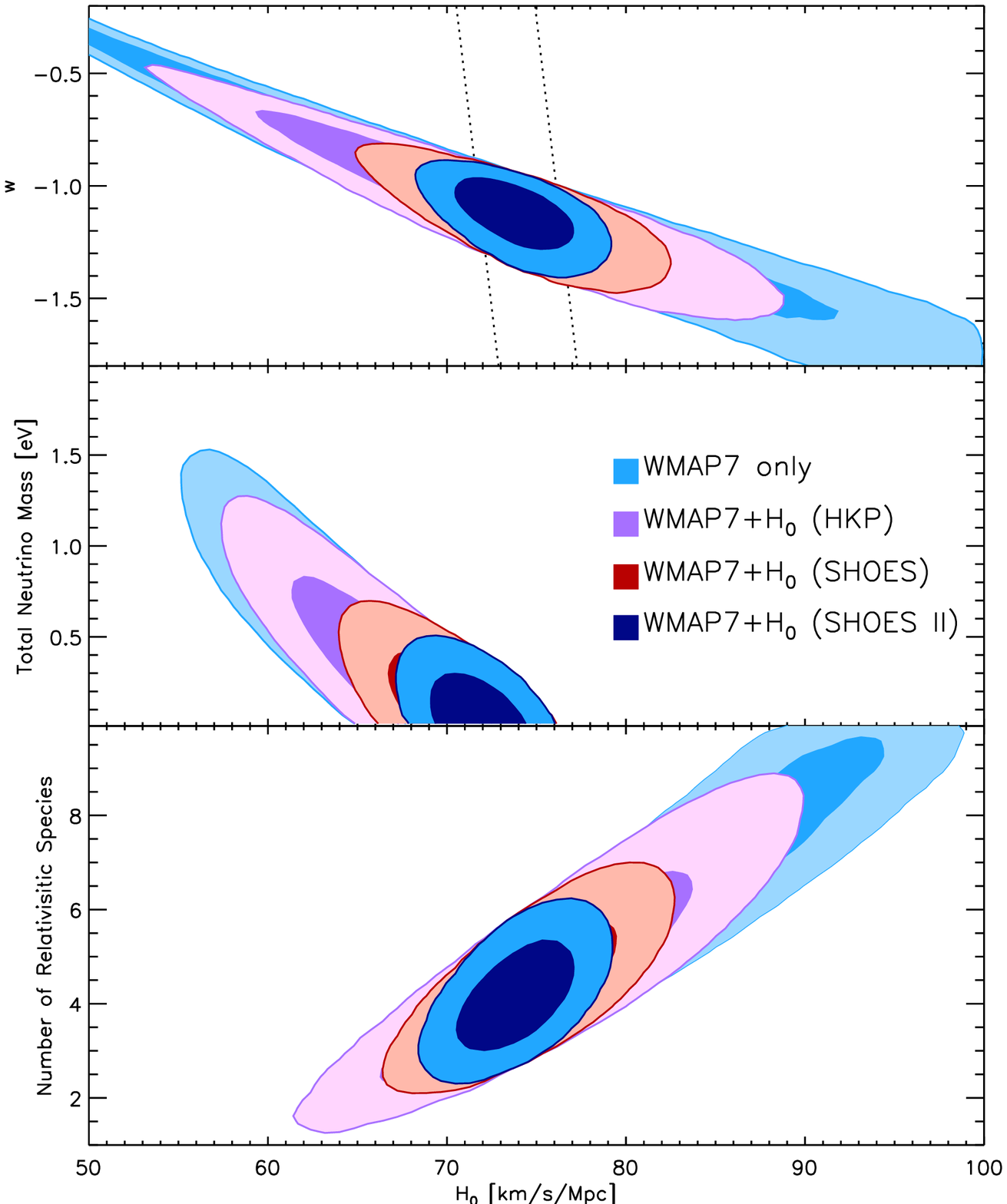}
\caption { }
\end{figure}

\end{document}